**Coordination Engineering of Dual-Atom Catalysts for Overall Water Splitting: Mechanistic Insights from Constant-Potential First-Principles and Machine Learning**

Jiahang Li[1], Suhang Li[1], Chong Yan[1], Jiajun Yu[1], Qinzhuang Liu[1,2], Ruo-Ya Wang[1*], Dongwei Ma[1*]

[1]Anhui Province Key Laboratory of Intelligent Computing and Applications, College of Physics and Electronic Engineering, Huaibei Normal University, Huaibei 235000, China

[2]Huaibei Key Laboratory for Advanced Thin Film Materials and Technology, Huaibei Normal University, Huaibei 235000, China

**Abstract**

The rational design of bifunctional electrocatalysts for the hydrogen evolution reaction (HER) and oxygen evolution reaction (OER) is essential for achieving efficient and cost-effective overall water splitting. Atomically dispersed transition-metal catalysts, including single-atom catalysts and dual-atom catalysts (DACs), have emerged as a prominent class of heterogeneous catalysts, in which coordination engineering plays a decisive role in tuning catalytic performance. Herein, we explore coordination-engineered bifunctional overall water splitting electrocatalysts using graphene-supported DACs ($TM_1TM_2$-$C_{6-x}N_x$) as model systems. By tuning C/N coordination and dual-metal combinations (Fe, Co, Ni, and Cu), a library of 228 structures was constructed. A three-step screening strategy, combining constant-charge and constant-potential density functional theory with kinetic analysis of proton-coupled electron transfer (PCET), identifies 24 highly active candidates ($TM_1TM_2$ = CoNi, CoCu and $Co_2$) with mixed C/N coordination for OER. These catalysts exhibit overpotentials comparable to that of $IrO_2$ and low PCET barriers (lower than 0.40 eV), among which 22 also show high HER activity. Machine learning reveals clear coordination-dependent

*Corresponding author. E-mail: WRYwangruoya@163.com (R. Wang)
*Corresponding author. E-mail: madw@chnu.edu.cn, dwmachina@126.com (D. Ma)

structure–performance relationships. Such bifunctionality arises from coordination engineering that enables the simultaneous optimization of OER intermediate adsorption and the hydrogen binding strength for HER. This work establishes coordination engineering as an effective strategy for designing high-performance bifunctional dual-atom electrocatalysts for overall water splitting.

**KEYWORDS:** *overall water splitting, coordination engineering, dual-atom catalysts, bifunctional electrocatalysts, machine learning, DFT*

## 1. Introduction

Hydrogen energy is a key enabler for bridging renewable and conventional energy systems, and its large-scale production is central to a carbon-neutral future.[1-4] Among various technologies, acidic overall water splitting—featuring structural simplicity, rapid proton transport, and compatibility with high current densities—represents a promising route for renewable hydrogen production.[5-7] This process involves the hydrogen evolution reaction (HER) at the cathode and the oxygen evolution reaction (OER) at the anode,[8,9] which typically rely on scarce noble-metal catalysts such as Pt and Ir/Ru oxides, respectively.[10-13] However, the use of separate single-function catalysts increases system complexity and cost, while hindering the synergistic optimization of HER and OER. Therefore, the development of efficient and low-cost bifunctional electrocatalysts capable of driving both reactions is highly desirable to simplify device architectures and improve overall performance.[14-18]

Reducing metal loading to achieve atomically dispersed catalysts represents an effective strategy for efficient and economical electrocatalysis.[19-21] In this context, graphene-based single-atom catalysts (SACs) and dual-atom catalysts (DACs) have attracted considerable attention owing to their tunable structures, high conductivity, and low cost.[22-25] Coordination engineering has emerged as a powerful approach to regulate the electronic structure and catalytic performance of these systems by strengthening metal–support interactions, modulating active-site electronic structures, and optimizing intermediate adsorption.[26-30] For example, Yan et al.[31] reported a FeCo-$N_3O_3$@C DAC with asymmetric N/O coordination, enabling optimized oxygen intermediate adsorption and efficient bifunctional OER/ORR activity. Li et al.[32] engineered an asymmetric $N_3$-Fe-Co-$N_4$ structure via Co incorporation, which promotes charge transfer and increases the density of states near the Fermi level, thereby enhancing OER/ORR performance. Liu et al.[33] demonstrated that coordination tuning of dual-atom sites optimizes key intermediate adsorption, affording Cu/Ni-$N_2C_2$ with excellent $CO_2$RR/OER bifunctionality.

Notably, the integration of atomic site anchoring with precise coordination regulation can further enhance both catalytic activity and stability under acidic HER and OER conditions. For instance, Zhao et al.[34] developed a PtRu DAC embedded into the N-doped carbon via a robust anchoring strategy, delivering superior HER activity compared to commercial 20 wt% Pt/C along with excellent stability in acidic media. The Lee group[35] synthesized a NiCo DAC on N-doped carbon through in situ metal ion capture coupled with coordination control, achieving high HER activity over a wide pH range. Kim et al.[36] reported Co–$N_4$ SACs with tailored nitrogen coordination, in which pyrrolic N sites facilitate intermediate adsorption and accelerate OER kinetics, resulting in enhanced activity and stability under acidic conditions. In addition, dual-site Ru/Co–$N_4$ catalysts exhibit efficient bifunctional OER/HER performance in proton exchange membrane electrolyzers.[37] Collectively, these studies underscore the critical role of structural regulation in enabling high-performance electrocatalysis for acidic OER and HER.

Inspired by these advances, we systematically investigate coordination-engineered bifunctional dual-atom electrocatalysts for acidic overall water splitting by integrating density functional theory (DFT), machine learning (ML), constant-potential simulations, and proton-coupled electron transfer (PCET) kinetics. A structural library of 228 $TM_1TM_2$-$C_{6-x}N_x$ DACs (TM = Fe, Co, Ni, and Cu) is constructed by tuning N/C coordination and dual-metal combinations. High-throughput screening identifies 24 candidates with low thermodynamic overpotentials (< 0.30 V) and PCET barriers (< 0.40 eV), among which 22 also exhibit high HER activity. Machine learning (ML) further reveals key coordination-dependent descriptors governing catalytic performance. These results demonstrate that tuning N/C coordination enables the simultaneous optimization of OER and HER activity in graphene-based DACs, thereby providing a coordination-engineering-based strategy for designing high-performance bifunctional electrocatalysts for overall water splitting.

## 2. Computational Method

All spin-polarized density functional theory (DFT) calculations were performed using the Vienna *ab initio* Simulation Package (VASP),[38] with the projector augmented wave (PAW) method employed to describe electron-ion interactions.[39] The Perdew–Burke–Ernzerhof (PBE) functional within the generalized gradient approximation (GGA) was adopted to treat exchange-correlation effects,[40] and the DFT-D3 scheme was used to correct van der Waals interactions.[41] The thermodynamic behaviors of OER and HER were explored using the computational hydrogen electrode (CHE) model,[42] Solvation effects were considered via the implicit solvation model implemented in VASPsol.[43]

The electrode potential was incorporated using the constant-potential (CP) method developed by Duan and Xiao.[44,45] Geometry optimizations were carried out within a grand canonical framework, wherein structures were optimized at a fixed potential by iteratively adjusting the electron count to align the Fermi level with the target potential. Kinetic analysis of PCET was carried out using the electrochemical nudged elastic band (eNEB) method.[46] Six intermediate images were interpolated between fully relaxed initial and final states to map the minimum energy pathway (MEP), and three $H_2O$ and one $H_3O^+$ were introduced to mimic acidic environments. The exchange current density ($i_0$) for HER was estimated from the hydrogen adsorption free energy ($\Delta G_{*H}$) using the Nørskov model, which correlates $\Delta G_{*H}$ with intrinsic proton-transfer kinetics.[47]

ML analyses were performed using Bayesian optimization-random forest (BO-RF) implemented in Scikit-learn[48-50] (4:1 train–test split; $R^2$ and RMSE as evaluation metrics) and the Sure Independence Screening and Sparsifying Operator (SISSO)[51] algorithm. This algorithm integrates symbolic regression and compressed sensing to screen physically interpretable low-dimensional descriptors, thereby elucidating catalyst structure–activity relationships. More computational details are available in the Supporting Information.

## 3. Results and Discussion

### 3.1. Structure of DACs and Workflow

As shown in Fig. 1a, a structural model library of $TM_1TM_2$-$C_{6-x}N_x$ DACs was constructed based on 3d transition metals (Fe, Co, Ni, and Cu). The metal pairs were embedded in graphene by occupying four adjacent carbon vacancies, with the vacancy edge atoms consisting of carbon atoms, nitrogen atoms, or mixed C/N species. Both homonuclear (including $Co_2$ and $Fe_2$) and heteronuclear (including CoCu, CoNi, FeCu, FeNi, and FeCo) metal pairs were considered. The coordination environment is described as $C_{6-x}N_x$, comprising seven distinct categories ranging from pure carbon ($C_6$) to full nitrogen coordination ($N_6$), namely $C_6$, $C_5N$, $C_4N_2$, $C_3N_3$, $C_2N_4$, $CN_5$, and $N_6$. Within each category, multiple configurations may arise depending on the specific spatial arrangement of C and N atoms. The number of such configurations varies significantly with coordination: $C_6$ and $N_6$ each have one configuration; $C_5N$ and $CN_5$ have three configurations for heteronuclear systems and two for homonuclear systems; $C_4N_2$ and $C_2N_4$ each have nine and six configurations, respectively; and $C_3N_3$ has ten configurations for heteronuclear systems and six for homonuclear systems. In total, 228 DAC structures were generated, with configuration indexing following the site-labeling scheme defined in Fig. S1. Note that among these configurations, FeCu-$N_6$,[52] $Cu_2$-$N_6$,[52] and FeNi-$N_6$[53] have been experimentally synthesized.

Fig. 1b outlines the overall workflow from model construction to performance evaluation. First, coordination engineering combined with diverse dual-metal configurations is employed to expand the chemical space of candidate catalysts. Subsequently, constant-charge DFT (CC-DFT) calculations are conducted to obtain the adsorption free energies of key OER intermediates ($\Delta G_{*OH}$, $\Delta G_{*O}$, and $\Delta G_{*OOH}$) as well as the OER overpotential ($\eta_{OER}$), enabling a preliminary assessment of catalytic performance. On this basis, machine learning models (including BO–RF and SISSO) are developed to uncover intrinsic structure–performance relationships for OER. Furthermore, constant-potential DFT (CP-DFT) calculations are performed to evaluate the overpotential under more realistic electrochemical conditions. Moreover, the activation barrier ($E_a$) of PCET process in the potential-determining step (PDS) is evaluated with explicit solvent molecules and potential effects taken into account. For HER, the $\Delta G_{*H}$ is calculated, and candidates with $\Delta G_{*H}$ close to the thermodynamic optimum (≈ 0) are selected. Overall, a multi-criteria evaluation strategy is employed to identify bifunctional electrocatalysts with outstanding OER and HER performance, thereby establishing an integrated workflow from high-throughput computation to rational catalyst design.

**3.2. Electrocatalytic Activity of OER**

We first evaluate the OER performance of $TM_1TM_2$-$C_{6-x}N_x$ DACs. Based on CC-DFT calculations, the adsorption free energies of key OER intermediates (*OH, *O, and *OOH) on 228 $TM_1TM_2$-$C_{6-x}N_x$ DACs were systematically computed (Tables S2–S8), from which $\eta_{OER}$ was derived. It is found that OER intermediates preferentially adsorb on the $TM_1$ top site, which possesses fewer d valence electrons, rather than on $TM_2$ or C/N sites, as illustrated in the inset of Fig. 2a. Fig. S2 presents the free energy diagrams of the most active configurations for each of the seven dual-metal combinations. Notably, a lower $\eta_{OER}$ corresponds to higher catalytic activity.[54-56] Figure 2a shows the distribution of $\eta_{OER}$ (0.05–1.55 V) of all systems using a color gradient, where cyan denotes low overpotentials and dark brown indicates high overpotentials. To benchmark against state-of-the-art $IrO_2$ catalysts, a screening threshold of $\eta_{OER} <$ 0.40 V was adopted.[57-59] Based on this criterion, 69 promising candidates were identified from the 228 structures.

Metal composition analysis shows that, among homonuclear systems, $Co_2$ exhibits outstanding OER activity, with most configurations meeting the high-activity criterion. In contrast, $Fe_2$ systems generally display higher $\eta_{OER}$ values, with only a few configurations approaching the threshold, indicating inferior catalytic performance. Among heteronuclear systems, CoCu and CoNi show activities comparable to $Co_2$, with most configurations satisfying $\eta_{OER} < 0.40$ V, whereas FeCu, FeNi, and FeCo systems exhibit consistently higher $\eta_{OER}$ values. In terms of coordination environment, highly active configurations ($\eta_{OER} < 0.40$ V) are predominantly associated with moderate nitrogen coordination (e.g., $C_2N_4$, $C_3N_3$, and $C_4N_2$), whereas purely carbon-coordinated ($C_6$) or highly nitrogen-coordinated structures (e.g., $CN_5$ and $N_6$) tend to exhibit elevated overpotentials. These results indicate that both the identity of the dual-metal centers and an optimal C/N coordination ratio are crucial for achieving high OER performance.

The catalytic activity is largely governed by the adsorption free energies of reaction intermediates. Previous studies have shown that scaling relationships among these intermediates can significantly accelerate the identification of high-performance catalysts.[60,61] As summarized in Tables S9–S15, the PDS for most structures is *O → *OOH, while for a smaller subset it is *OH → *O; in both cases, the *O intermediate plays a central role. The relationships among $\Delta G_{*OH}$, $\Delta G_{*O}$, and $\Delta G_{*OOH}$ are shown in Figs. 2b, 2c and S3. Strong linear correlations are observed between $\Delta G_{*O}$ and $\Delta G_{*OH}$, as well as between $\Delta G_{*O}$ and $\Delta G_{*OOH}$, with coefficients of determination ($R^2$) of 0.913 and 0.943, respectively. Therefore, $\Delta G_{*O}$ can be employed as a reliable activity descriptor for $TM_1TM_2\text{-}C_{6\text{-}x}N_x$ DACs. The results further reveal a volcano-type relationship between $\eta_{OER}$ and $\Delta G_{*O}$, with the optimal OER performance achieved at $\Delta G_{*O} \approx 2.60$ eV (Fig. 2d).

### 3.3. ML Investigation

In the following, ML-based data mining was performed to gain fundamental insights into the relationship between structure and OER activity. Here, $\eta_{OER}$ was defined as the target variable, and 18 descriptors spanning elemental properties, electronic characteristics, and coordination features (Table S16) were initially considered. Although a larger feature set may improve predictive performance, it can reduce training efficiency and introduce redundancy or bias. Therefore, a reduced subset of descriptors was selected via correlation analysis.[62-65] As shown in Fig. 3a, the Pearson correlation coefficient ($p$) matrix was used to quantify the linear correlations among descriptors. The color gradient reflects both the sign and magnitude of correlations (red for positive, blue for negative, with intensity corresponding to $|p|$). Strong correlations are observed between the C and N coordination numbers of $TM_1$ ($N_{C1}$ and $N_{N1}$) as well as $TM_2$ ($N_{C2}$ and $N_{N2}$). To minimize multicollinearity, 16 descriptors with $|p| < 0.9$ were retained for subsequent modeling, while $N_{N1}$ and $N_{N2}$ were excluded.

The dataset derived from DFT calculations on 228 DAC structures was randomly divided into a training set (80%) and an independent test set (20%). The BO–RF model shows excellent agreement with DFT results (Fig. 3b and Table S17), achieving an $R^2$ of 0.91, an RMSE of 0.08 V, and a mean absolute error (MAE) of 0.06 V on the test set, indicating high predictive reliability. The feature importance analysis (Fig. 3c) reveals that $Q_1$, $EN_1$–$EN_2$, and $N_{d1}$+$N_{d2}$ are the most influential descriptors, contributing 35.56%, 20.88%, and 10.59%, respectively. Here, $Q_1$ is the net charge of $TM_1$, $EN_1$–$EN_2$ represents the electronegativity difference between the two metal centers, and $N_{d1}$+$N_{d2}$ denotes the total number of d valence electrons. Notably, $Q_1$ exhibits the highest importance, highlighting the critical role of charge-transfer characteristics in determining OER activity. Other descriptors, including $M_1$, $Q_2$, $\Delta EN_1$, $Nd_1$–$Nd_2$, $NC_1$, $\Delta EN_2$, and $N_{C2}$, show comparatively minor contributions. Overall, $\eta_{OER}$ is primarily governed by the electronic structure of the active sites and their coordination environments.

To further extract physically meaningful insights, we constructed an optimal composite descriptor by combining the most relevant features using the SISSO approach.[51] This method establishes an explicit and interpretable relationship between key descriptors and $\eta_{OER}$. The resulting two-dimensional descriptor with the highest relevance is expressed as follows:

$$\eta_{\mathrm{OER}} = 42.646\frac{\mathrm{Q_1}}{(\mathrm{N_{d1}} + \mathrm{N_{d2}})(\Delta\mathrm{EN_1} + \Delta\mathrm{EN_2})} + 0.125\sqrt{\Delta\mathrm{EN_2}} * (\mathrm{N_{d1}} + \mathrm{N_{d2}}) - 3.328$$

As described above, $Nd_1 + Nd_2$ represents the total number of valence *d*-electrons of the dual-metal sites, while $\Delta EN_1$ and $\Delta EN_2$ denote the electronegativity differences between each metal center and the average electronegativity of its coordinated nonmetal atoms. $Q_1$ reflects the charge-transfer propensity of $TM_1$. The first term describes the interplay between charge-transfer strength and the system's ability to accommodate charge redistribution. A larger $Q_1$ indicates a stronger tendency of $TM_1$ to donate or accept electrons, facilitating modulation of the electronic structure of reaction intermediates. In contrast, an increased total *d*-electron count or larger electronegativity difference enhances the capacity for charge redistribution, thereby mitigating excessive charge accumulation and its unfavorable impact on the overpotential. The second term is primarily associated with the coordination environment of $TM_2$. A larger $\Delta EN_2$, corresponding to a greater electronegativity mismatch between $TM_2$ and its coordinating atoms, leads to a stronger contribution from this term, which is further amplified by an increased total *d*-electron count. This term reflects the role of $TM_2$ and its coordination environment in tuning the electronic structure and thereby influencing the stability of reaction intermediates.

Overall, the OER activity is governed by the cooperative effects of both metal centers and their coordination environments, where charge-transfer characteristics and coordination-dependent electronic modulation jointly determine catalytic performance. As shown in Fig. 3d, the SISSO-predicted overpotentials correlate well with DFT results ($R^2$ = 0.88, RMSE = 0.10 V), demonstrating that the derived descriptor captures the key factors governing OER activity. This level of interpretability provides valuable insights into structure–property relationships and supports the rational design of high-performance catalysts.

### 3.4. Constant-Potential Simulation of OER

All the above OER thermodynamic calculations were performed under constant-charge conditions within the CHE framework, which effectively captures activity trends across a large set of catalysts; however, this approach inherently assumes an overall charge-neutral system.[66] In practical electrochemical environments, charge exchange between the catalyst and electrode occurs continuously while the electrode potential is maintained constant, necessitating a constant-potential description.[67-69] To bridge this gap, CP-DFT calculations were carried out at 0.00 $V_{RHE}$ and pH = 1 for the 69 $TM_1TM_2$-$C_{6-x}N_x$ DACs ($TM_1TM_2$ = CoNi, CoCu and $Co_2$) identified in the initial screening. Adsorption free energies are summarized in Tables S18–S20, while the corresponding reaction free energies, PDS, and overpotentials are listed in Tables S21–S23. The resulting $\eta_{OER}$ values are presented in Fig. 4a. Notably, the overpotentials predicted by the constant-potential model are in good agreement with those obtained from the constant-charge calculations, validating the reliability of the initial high-throughput screening while providing a more realistic electrochemical description.

To further identify the most competitive candidates, a more stringent criterion of $\eta_{OER} < 0.30$ V was applied, yielding 24 highly active OER catalysts. To evaluate their catalytic behavior under operating conditions, we further examined the reaction free energies at $U_{RHE} = \Delta G_{max}/e$ and investigated the kinetics of the PCET of PDS.[70] Explicit water molecules were incorporated to model the acidic environment, where PCET proceeds via proton transfer from adsorbed intermediates to solvent water, forming $H_3O^+$ species. Six intermediate images were interpolated between the initial and final states to determine the minimum energy pathway. The corresponding free energy profiles and reaction pathways are shown in Figs. 4b and S4–S6.

Among the 24 candidates, 12 exhibit *OH → *O as the PDS, whereas the remaining 12 follow *O → *OOH, with the *O intermediate playing a central role in both pathways. Taking the CoNi system as an example (Fig. 4b), the PDS for $C_3N_3$-124 corresponds to *OH → *O, involving proton transfer from *OH to a solvent water molecule and the formation of a stabilized hydronium cluster ($H_3O^+ + 3H_2O$), with an activation barrier of 0.10 eV. In contrast, for $C_5N$-2, the PDS is *O → *OOH, where O–O bond formation occurs via interaction between *O and a water molecule, coupled with proton transfer, yielding *OOH with a barrier of 0.28 eV.

The PDS energy barriers of the 24 candidates are summarized in Fig. 4c. Notably, all systems exhibit low kinetic barriers ($E_a < 0.40$ eV), indicating rapid OER kinetics. These results highlight the critical role of explicit solvent participation in facilitating proton transfer and lowering kinetic barriers under acidic conditions. Therefore, from both thermodynamic and kinetic perspectives, the identified 24 $TM_1TM_2$-$C_{6-x}N_x$ DACs exhibit outstanding OER activity. Despite their excellent OER performance, the ultimate goal of this work is to determine whether coordination engineering can endow these systems with bifunctional electrocatalytic activity. Accordingly, in addition to anodic OER activity, the cathodic HER performance of these candidates is further evaluated as a key criterion for overall water splitting.

**3.5. Electrocatalytic Activity of HER**

An ideal HER catalyst under acidic conditions typically exhibits a $\Delta G_{*H}$ close to 0.0 eV, corresponding to an optimal hydrogen binding strength.[71] For comparison, CoNi, CoCu, and $Co_2$ DACs with $N_6$ coordination were first investigated at 0.00 $V_{RHE}$, considering both TM sites and three inequivalent N sites. The calculated $\Delta G_{*H}$ values (Table S24) indicate that both TM and N sites exhibit weak hydrogen binding, with $\Delta G_{*H}$ values (0.47–1.94 eV) significantly larger than the thermoneutral value (~0 eV), rendering them unfavorable for the Volmer step of HER. Among these, the most favorable adsorption occurs at one of the N sites, although its $\Delta G_{*H}$ remains far from the optimal range.

In contrast, for the 24 candidates with mixed C/N coordination (Tables S25), the $\Delta G_{*H}$ values decrease markedly, indicating a substantial modulation of hydrogen binding strength toward the optimal range. In some configurations, $\Delta G_{*H}$ even becomes negative (e.g., −0.20 to −0.29 eV), suggesting a transition from weak to moderate or slightly strong adsorption. These results demonstrate that coordination engineering effectively tunes the interaction between active sites and hydrogen intermediates, enabling optimization of HER activity. Structural optimization results (Fig. S7) further reveal that, when Co atoms are coordinated with carbon, hydrogen preferentially adsorbs at the Co–C bridge site. In contrast, for $TM_2$ sites (TM = Ni, Cu), hydrogen adsorption favors adjacent carbon sites rather than TM–C bridge positions. Moreover, in most cases, hydrogen initially adsorbed on TM sites tends to migrate to more stable configurations, namely Co–C bridge sites (for Co site) or carbon sites (for Ni and Cu sites). These distinct adsorption behaviors reflect the sensitivity of hydrogen binding to the local coordination environment, thereby providing a structural basis for tuning HER activity.

The $\Delta G_{*H}$ values for $TM_1TM_2$-$C_{6-x}N_x$ DACs are more intuitively summarized in Fig. 5a. The shaded pink region (−0.2 to +0.2 eV) highlights promising HER candidates with activities comparable to Pt.[72,73] To further quantify catalytic performance, the exchange current density ($i_0$) was plotted as a function of $\Delta G_{*H}$ (Fig. 5b), constructing a volcano-type relationships based on the Nørskov model.[47]

$$i_0 = \frac{-ek_0}{1 + \exp\left(|\Delta G_{*\mathrm{H}}/k_b T|\right)}$$

All three systems ($Co_2$, CoNi, and CoCu) exhibit the characteristic volcano behavior consistent with the Sabatier principle,[74] where $\log(i_0)$ reaches its maximum at $\Delta G_{*H} \approx 0.00$ eV, corresponding to optimal hydrogen binding strength and thus enhanced HER activity. Notably, the apex of the volcano closely aligns with the optimal $\Delta G_{*H}$ window identified in Fig. 5a, further validating $\Delta G_{*H}$ as a reliable activity descriptor for $TM_1TM_2$-$C_{6-x}N_x$ DACs. Based on these results, the 22 $TM_1TM_2$-$C_{6-x}N_x$ DACs with mixed C/N coordination exhibit excellent catalytic activity for both OER and HER, demonstrating their strong potential as bifunctional electrocatalysts.

**3.6. Electronic Structure Analysis**

To elucidate the influence of the coordination environment on the bifunctional overall water-splitting performance of $TM_1TM_2$-$C_{6-x}N_x$ DACs, $Co_2$-$N_6$ and $Co_2$-$C_4N_2$-25 were selected as representative models to comparatively investigate the performance evolution and underlying modulation mechanism during the transition from pure N coordination to mixed C/N coordination ($C_4N_2$). Specifically, the reaction pathways and electronic structure characteristics of the active sites during the OER process were systematically analyzed for these two configurations. For both structures, the PDS is identified as *OH → *O. As shown in Figs. 6a and 6b, the corresponding Gibbs free energy changes ($\Delta G$) for this step are 1.80 eV and 1.50 eV for $Co_2$-$N_6$ and $Co_2$-$C_4N_2$-25, respectively. The free energy diagrams indicate that $Co_2$-$C_4N_2$-25 exhibits stronger adsorption toward *OH and *O, with $\Delta G_{*OH}$ and $\Delta G_{*O}$ decreasing from 1.47 and 3.27 eV ($Co_2$-$N_6$) to 1.02 and 2.52 eV, respectively. Notably, $\Delta G_{*OH}$ is reduced by 0.45 eV, whereas $\Delta G_{*O}$ decreases more significantly by 0.75 eV. This asymmetric stabilization of reaction intermediates primarily accounts for the reduced free energy change of the PDS. Consequently, the theoretical overpotential of $Co_2$-$C_4N_2$-25 decreases from 0.57 to 0.27 V, indicating a more favorable thermodynamic driving force for OER.

To further probe the origin of this performance enhancement, the projected density of states (PDOS) of key intermediates (*OH and *O) and the corresponding Co 3*d* orbitals were analyzed. Compared with $Co_2$-$N_6$, $Co_2$-$C_4N_2$-25 exhibits stronger orbital hybridization between Co 3*d* states and the intermediates, with more pronounced overlap near the Fermi level. Consistently, crystal orbital Hamilton population (COHP) analysis shows that the integrated COHP (ICOHP) values become more negative upon coordination modulation, changing from −4.08 to −4.29 eV for *OH adsorption and from −6.67 to −7.35 eV for *O adsorption, in agreement with the enhanced adsorption strengths. In addition, as illustrated in Fig. S8, the d-band center ($\varepsilon_d$) of the Co atom in $Co_2$-$C_4N_2$-25 shifts upward from −0.86 eV to −0.48 eV relative to that in $Co_2$-$N_6$, approaching the Fermi level. This shift correlates with strengthened interactions between the active sites and oxygen-containing intermediates, thereby tuning the adsorption strength toward a more optimal regime and contributing to the improved OER catalytic activity.

During the HER process, coordination regulation likewise plays a critical role. As shown in Figs. 6c and S7, in the purely N-coordinated $Co_2$-$N_6$ system, *H preferentially adsorbs at N sites, with a $\Delta G_{*H}$ of 0.94 eV at the N1 site, indicating overly weak hydrogen binding. Upon introducing C coordination ($Co_2$-$C_4N_2$-25), the preferred adsorption site shifts to the Co–C bridge site, where $\Delta G_{*H}$ decreases to −0.14 eV. This value lies much closer to the thermoneutral condition, suggesting that the hydrogen binding strength is effectively tuned toward the optimal range. Further insights from PDOS and COHP analyses provide an electronic-structure basis for this modulation. In the $Co_2$-$N_6$ system, the hybridization between H 1s and N 2*p* orbitals results in a less favorable interaction (ICOHP = −7.44 eV) compared with that in the $Co_2$-$C_4N_2$-25 system. In the latter, the H atom is simultaneously bonded to carbon and cobalt atoms, and the H 1s orbital interacts with Co 3*d* and C 2*p* orbitals, leading to a more negative ICOHP value (−7.86 eV).

Overall, the transition from pure N coordination to mixed C/N coordination effectively modulates the electronic structure of the active centers, thereby tuning the hydrogen binding strength toward an optimal regime. Together with the improved adsorption characteristics of oxygenated intermediates in OER, this coordination-induced electronic modulation enables a synergistic enhancement of bifunctional catalytic performance. These results highlight coordination engineering as an effective strategy for the rational design of high-performance electrocatalysts for overall water splitting.

Although theoretical calculations indicate that 22 $TM_1TM_2$-$C_{6-x}N_x$ DACs possess promising potential as bifunctional electrocatalysts for overall water splitting, experimental reports on the precise regulation of coordination environments and the synthesis of such structures remain limited. As shown in Fig. S9, the formation energies of these 22 candidate configurations range from 5.32 to 8.84 eV. In comparison, the formation energies of purely N-coordinated systems fall within 2.75–4.69 eV (Fig. S10). Notably, these values are generally comparable to those of experimentally synthesized counterparts, such as CoIn-$N_6$,[75] $Cu_2$-$N_6$,[52] $N_4$-Cu-Se-$C_3$,[76] and MnZn-$N_4S_2$,[77] with formation energies of 5.25, 5.74, 9.91, and 10.63 eV, respectively. Furthermore, all candidate systems exhibit negative binding energies ($E_b$, −11.71 to −19.07 eV), indicating strong anchoring of the dual-metal sites on the support. Collectively, these results suggest that the proposed catalysts are experimentally feasible for synthesis.

**4. Conclusion**

In this work, a comprehensive computational framework integrating high-throughput DFT calculations, constant-potential simulations, kinetic analysis, and interpretable machine learning was employed to investigate $TM_1TM_2$-$C_{6-x}N_x$ DACs for acidic overall water splitting. Systematic screening of 228 configurations reveals that both metal composition and coordination environment critically determine catalytic performance, with moderate C/N coordination enabling optimal adsorption of oxygenated intermediates and thus lowering OER overpotentials. Constant-potential and PCET kinetic analyses further confirm that the identified candidates exhibit not only favorable thermodynamics but also low activation barriers ($E_a$ < 0.40 eV), indicative of fast OER kinetics under realistic conditions. Machine learning analysis identifies key descriptors governing activity and establishes an interpretable relationship between structural features and $\eta_{OER}$, highlighting the cooperative roles of dual-metal electronic structure and coordination-induced modulation. Importantly, coordination engineering also effectively tunes hydrogen adsorption behavior, shifting $\Delta G_{*H}$ toward thermoneutral values and enabling competitive HER activity. Mechanistic investigations reveal that the transition from pure N coordination to mixed C/N coordination optimizes the electronic structure of active sites, thereby simultaneously improving adsorption energetics for both OER and HER intermediates. As a result, 22 DACs are identified with outstanding bifunctional performance. Overall, this study demonstrates that coordination engineering provides a viable strategy to simultaneously regulate thermodynamics, kinetics, and electronic structure in DACs, offering a clear design principle for developing efficient bifunctional electrocatalysts for acidic overall water splitting.

**Acknowledgements**

This work was supported by the Key Natural Science Research Project for Colleges and Universities of Anhui Province (No. 2025AHGXZK20053), the Innovative Research Team (in Science and Technology) in University of Henan Province (Grant No. 25IRTSTHN015), and the Natural Science Foundation of Henan (Grant No. 252300421499).

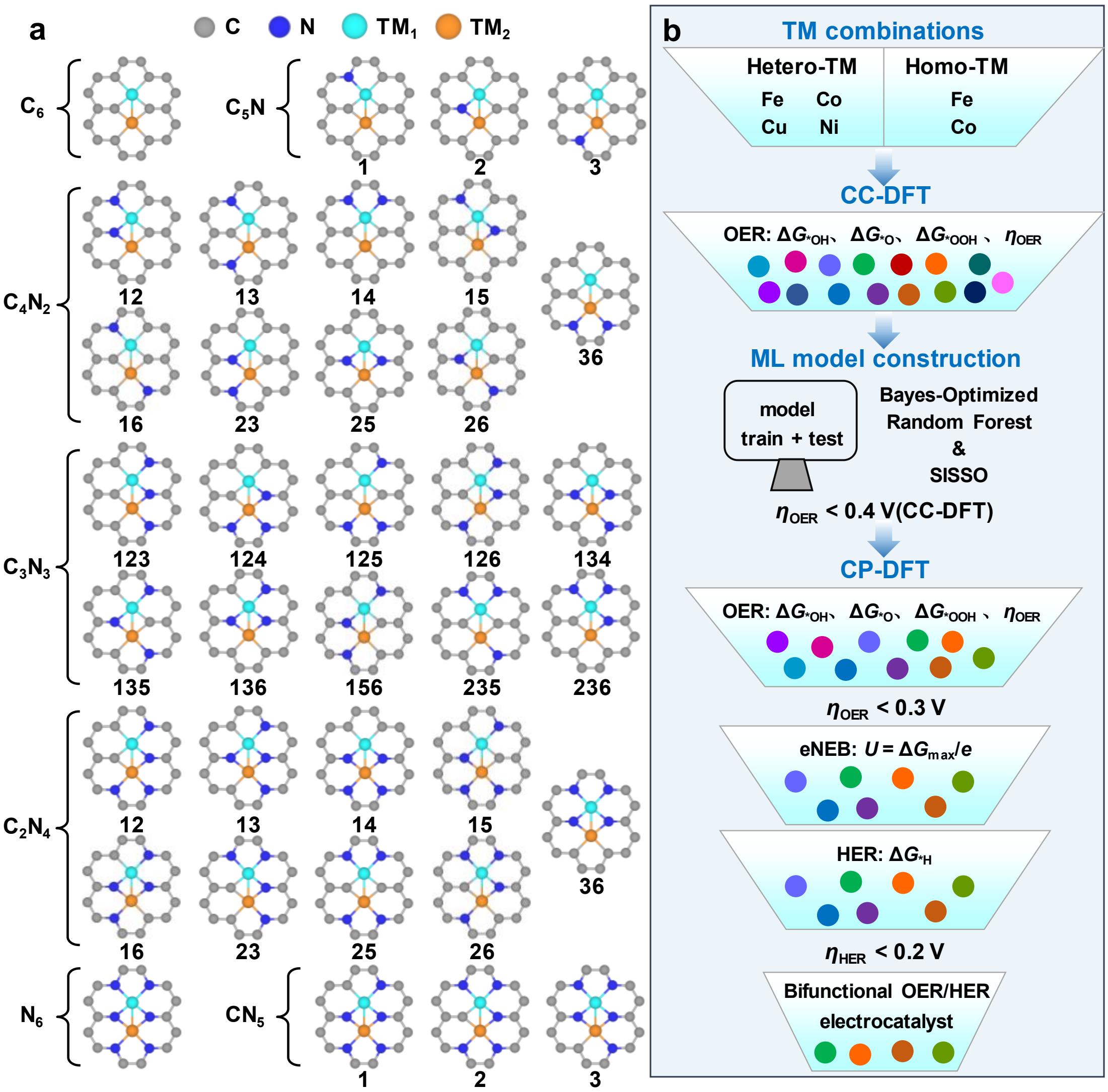

**Fig. 1.** (a) Structural models of $TM_1TM_2$-$C_{6-x}N_x$ DACs ($TM_1TM_2$ = $Co_2$, $Fe_2$, CoCu, CoNi, FeCu, FeNi, and FeCo). The coordination environments span $C_6$ to $N_6$ ($C_6$, $C_5N$, $C_4N_2$, $C_3N_3$, $C_2N_4$, $CN_5$, and $N_6$). A single configuration is identified for $C_6$ and $N_6$. For mixed C/N coordination environments, $C_5N$ and $CN_5$ exhibit three (heteronuclear) and two (homonuclear) configurations; $C_4N_2$ and $C_2N_4$ have nine and six configurations, respectively; and $C_3N_3$ has ten (heteronuclear) and six (homonuclear) configurations. Here, x denotes the number of N atoms; for $x < 3$, indices label N positions, whereas for $x \geq 3$, they label C positions (see Figure S1). (b) Workflow for the design and screening of bifunctional OER/HER electrocatalysts based on thermodynamic and kinetic criteria.

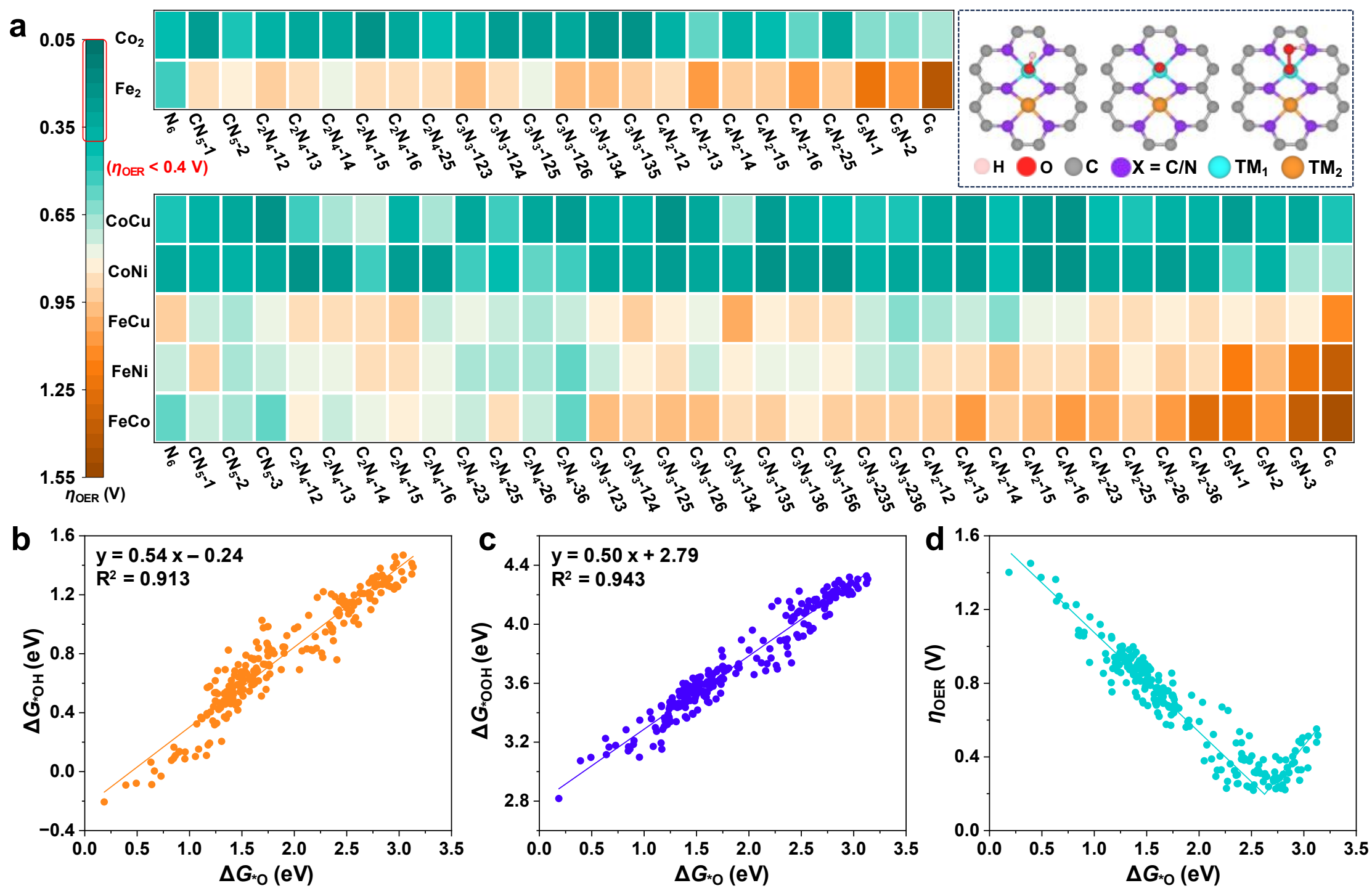

**Fig. 2.** (a) OER overpotentials ($\eta_{OER}$) of $TM_1TM_2$-$C_{6-x}N_x$ DACs, ($TM_1TM_2$ = $Co_2$, $Fe_2$, CoCu, CoNi, FeCu, FeNi, and FeCo). The upper-right inset shows representative adsorption configurations of key OER intermediates. Linear relationships of $\Delta G_{*OH}$ (b), $\Delta G_{*OOH}$ (c), and $\eta_{OER}$ (d) versus $\Delta G_{*O}$ for all $TM_1TM_2$-$C_{6-x}N_x$ DACs.

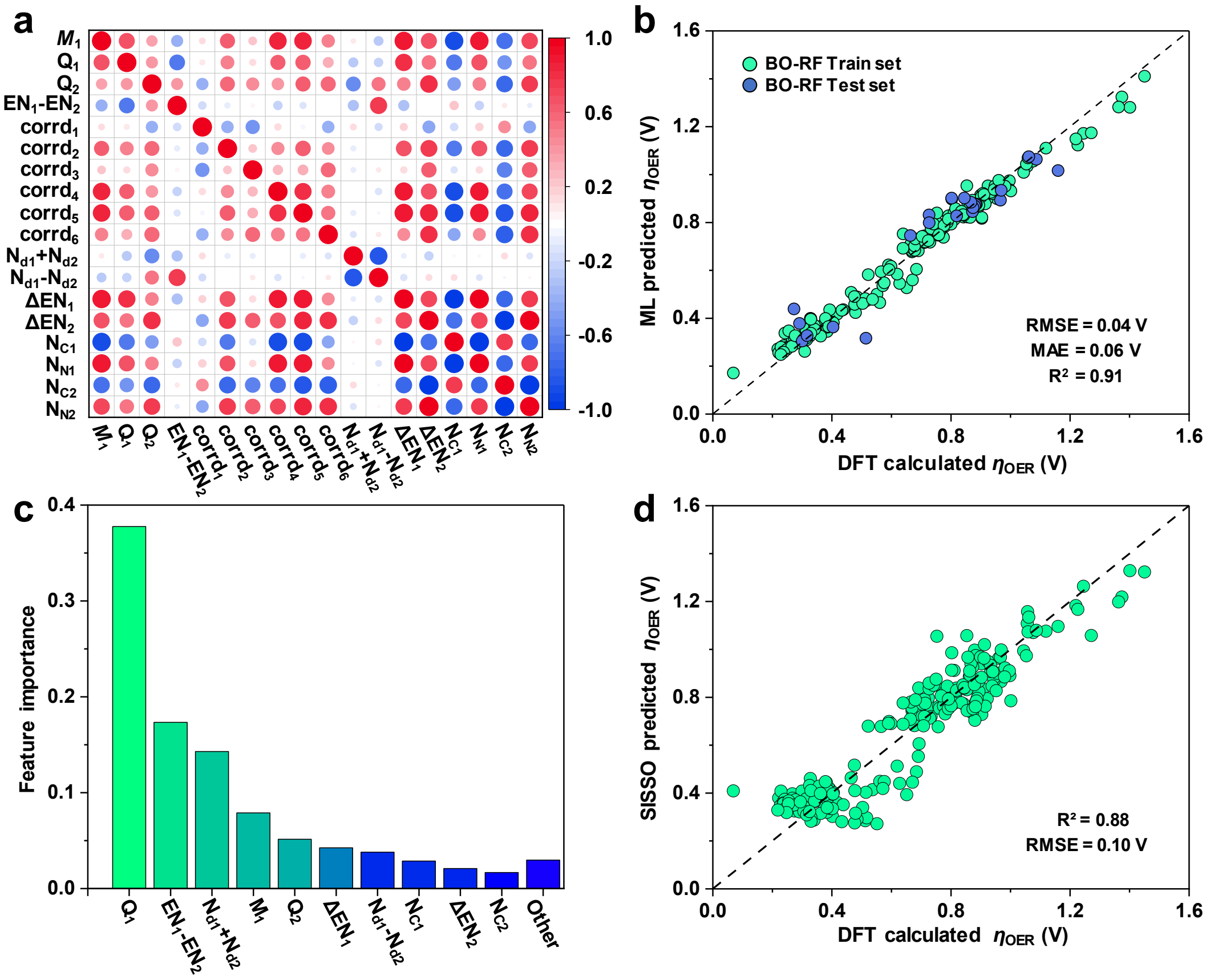

**Fig. 3.** (a) Pearson correlation heatmap of 18 features. (b) Comparison of DFT calculated and BO–RF predicted $\eta_{OER}$. (c) Feature importance of $\eta_{OER}$ calculated by the BO–RF model. (d) Comparison of DFT calculated and SISSO predicted $\eta_{OER}$.

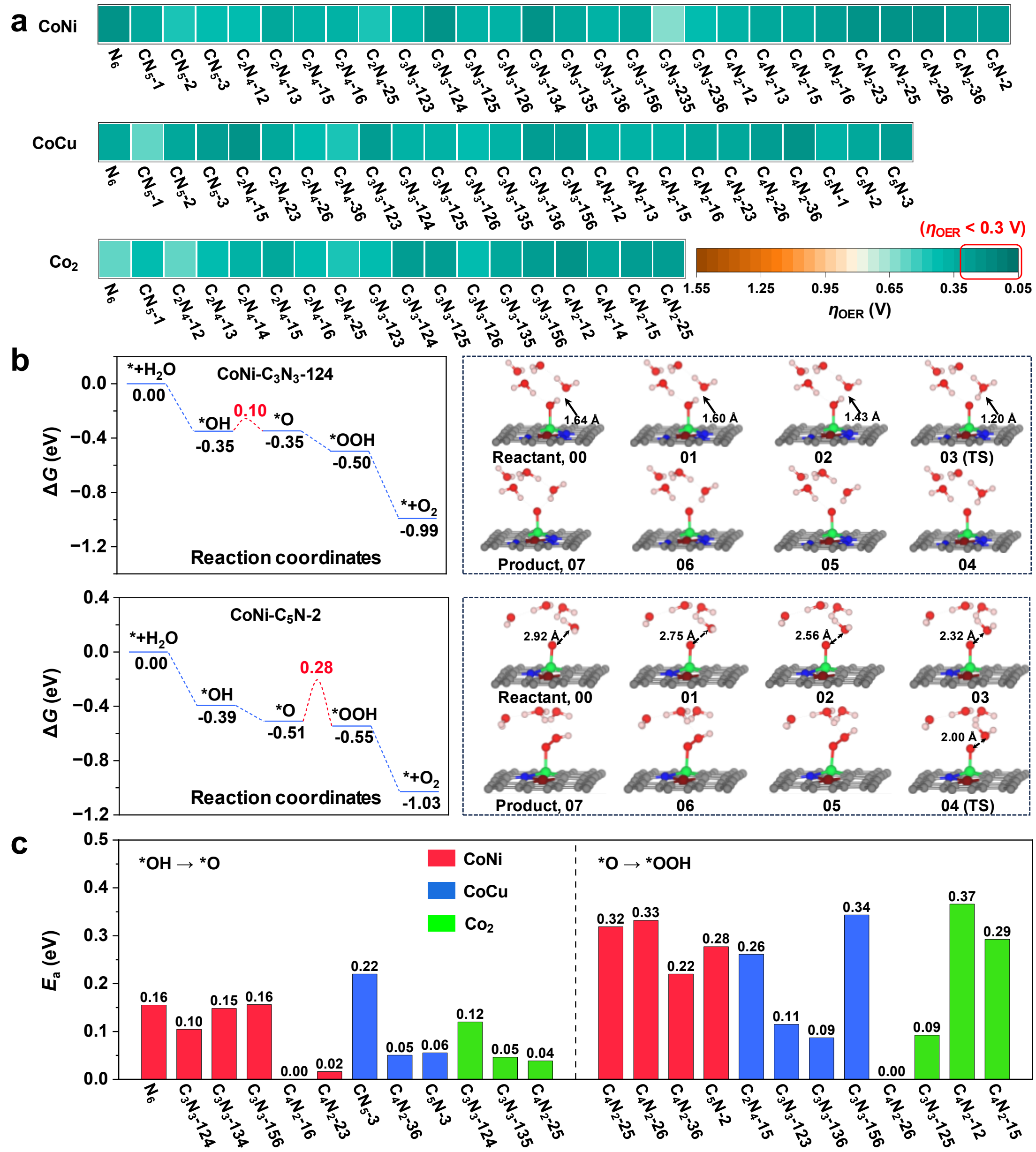

**Fig. 4.** (a) Calculated OER overpotentials ($\eta_{OER}$) of $TM_1TM_2$-$C_{6-x}N_x$ DACs ($TM_1TM_2$ = CoNi, CoCu and $Co_2$) at $U_{RHE}$ = 0.00 V. (b) Free energy diagrams for the OER on CoNi-$C_3N_3$-124 and CoNi-$C_5N$-2, along with the corresponding minimum energy paths for proton-coupled electron transfer (PCET) in the potential-determining steps (PDS), i.e., *OH → *O (upper) and *O → *OOH (lower), evaluated at $U_{RHE} = \Delta G_{max}/e$. The transition states (TS) correspond to image 03 (*OH → *O) and image 04 (*O → *OOH). (c) Summary of PCET energy barriers for the PDS of *O → *OOH (left) and *OH → *O (right) at $U_{RHE} = \Delta G_{max}/e$.

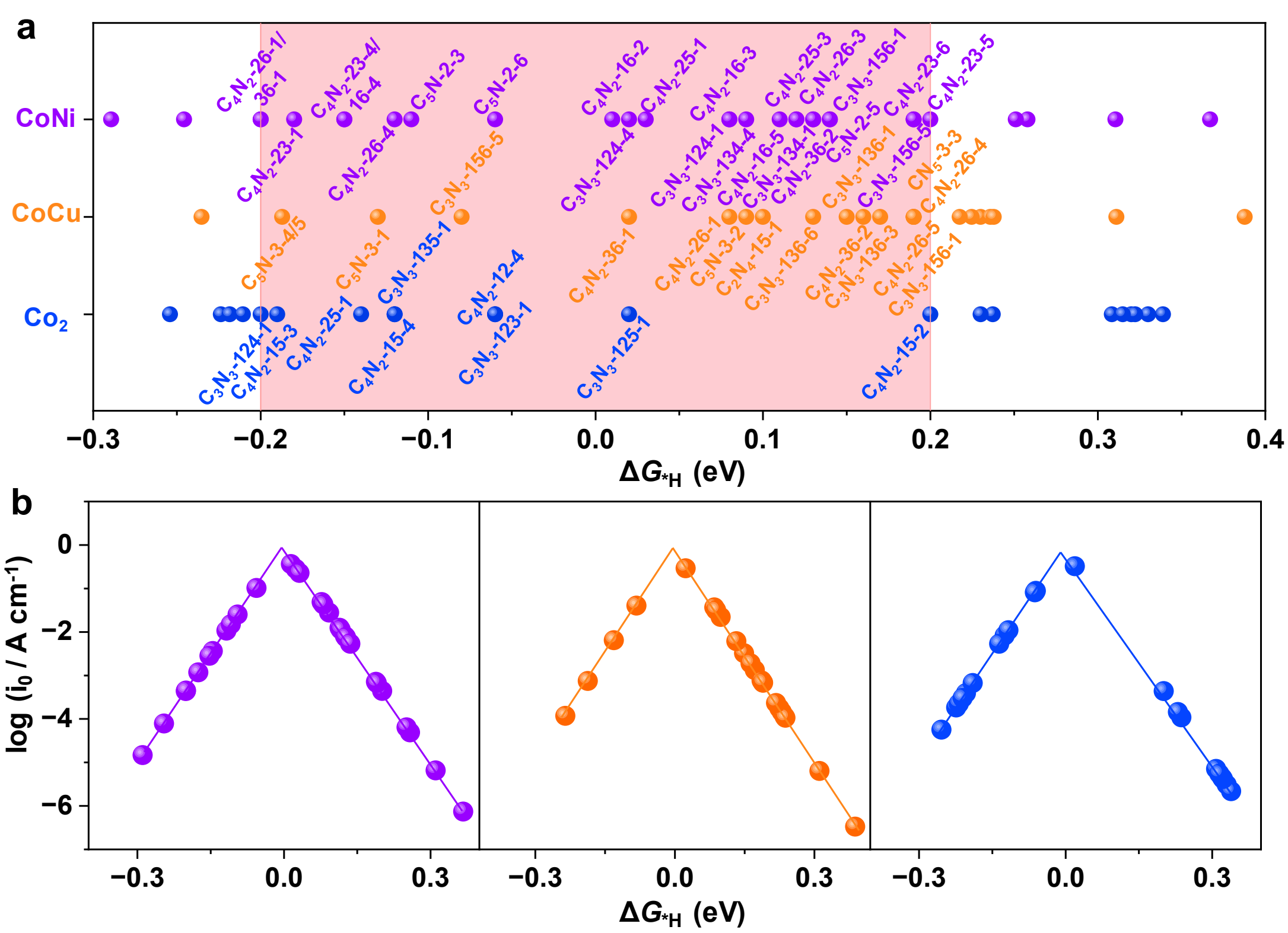

**Fig. 5.** (a) Hydrogen adsorption free energy ($\Delta G_{*H}$) of $TM_1TM_2$-$C_{6-x}N_x$ DACs with high OER activity ($TM_1TM_2$ = CoNi, CoCu, and $Co_2$) at $U_{RHE}$ = 0.00 V. The pink shaded region highlights near-optimal $\Delta G_{*H}$ window (between −0.2 and +0.2 eV). (b) Volcano plots of exchange current density ($i_0$) as a function of $\Delta G_{*H}$ for all adsorption sites in $TM_1TM_2$-$C_{6-x}N_x$ DACs ($TM_1TM_2$ = CoNi, CoCu and $Co_2$).

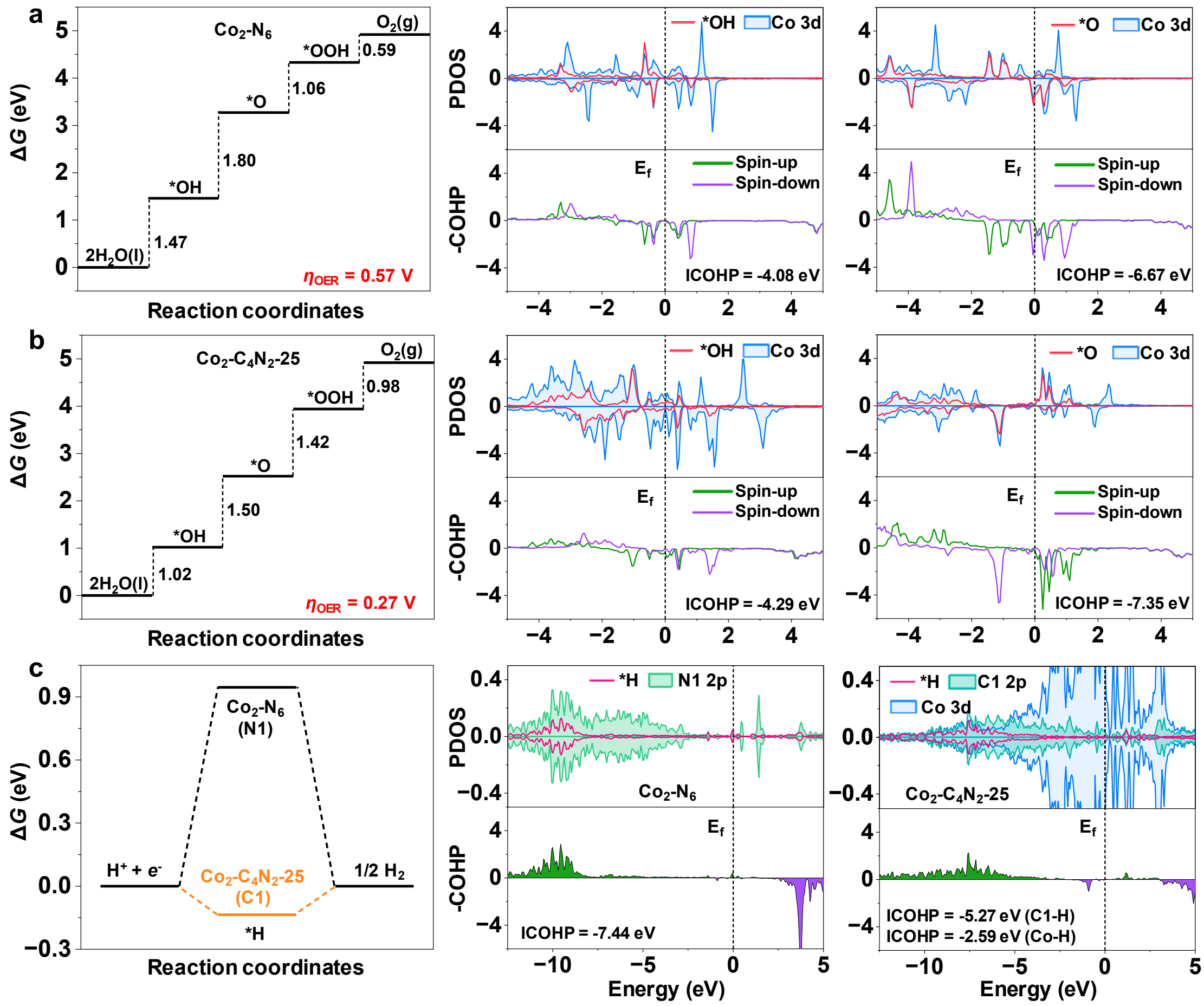

**Fig. 6.** (a) OER free energy diagram and electronic structure analysis of key intermediates on $Co_2$-$N_6$, including the free energy profile (left), projected density of states (PDOS) of *OH/*O and Co 3*d* orbitals, and crystal orbital Hamilton population (COHP) analysis of Co–O interactions (middle and right). (b) Same as in (a), but for $Co_2$-$C_4N_2$-25. (c) HER free energy diagrams of $Co_2$-$N_6$ and $Co_2$-$C_4N_2$-25 (left); PDOS of *H and N1 2*p* orbitals with COHP analysis of N–H interaction in $Co_2$-$N_6$ (middle); PDOS of *H, C1 2*p*, and Co 3*d* orbitals with COHP analyses of Co–H and C–H interactions in $Co_2$-$C_4N_2$-25 (right). All free energy calculations were performed under constant-potential conditions at 0.00 $V_{RHE}$. In the COHP plot, positive and negative values correspond to bonding and antibonding interactions, respectively. The Fermi level is set to 0 eV.

**Coordination Engineering of Dual-Atom Catalysts for Overall Water Splitting: Mechanistic Insights from Constant-Potential First-Principles and Machine Learning**

Jiahang Li[1], Suhang Li[1], Chong Yan[1], Jiajun Yu[1], Qinzhuang Liu[1,2], Ruo-Ya Wang[1*], Dongwei Ma[1*]

[1]Anhui Province Key Laboratory of Intelligent Computing and Applications, College of Physics and Electronic Engineering, Huaibei Normal University, Huaibei 235000, China

[2]Huaibei Key Laboratory for Advanced Thin Film Materials and Technology, Huaibei Normal University, Huaibei 235000, China

## Supporting Information

*Corresponding author. E-mail: WRYwangruoya@163.com (R. Wang)

*Corresponding author. E-mail: madw@chnu.edu.cn, dwmachina@126.com (D. Ma)

## Computational methods

All spin-polarized DFT calculations were carried out using the Vienna *ab initio* Simulation Package (VASP).[1] The interaction between valence electrons and ionic cores was described using the projector augmented wave (PAW) method.[2] The exchange-correlation effects were described within the generalized gradient approximation (GGA) using the Perdew-Burke-Ernzerhof (PBE) functional.[3] Van der Waals interactions were considered using the DFT-D3 correction scheme.[4] The plane-wave basis set was used with a cutoff energy of 400 eV. The convergence criteria were set to $10^{-5}$ eV for the total energy and 0.02 eV/Å for the Hellmann–Feynman forces. A $6 \times 3$ graphene supercell with an 18 Å vacuum layer was used to construct the $TM_1TM_2$-$C_{6-x}N_x$ DACs models. A $3 \times 2 \times 1$ Monkhorst–Pack k-point mesh was used for structural optimization, while a denser $7 \times 7 \times 1$ grid was employed for electronic structure calculations.[5]

The thermodynamics of the oxygen evolution reaction (OER) were investigated using the computational hydrogen electrode (CHE) model.[6] The reaction free energy of each elementary step was calculated according to $\Delta G = \Delta E + \Delta E_{ZPE} - T\Delta S - eU_{RHE}$, where $\Delta E$ is the reaction energy obtained from DFT calculations, $\Delta E_{ZPE}$ and $T\Delta S$ represents the contributions from the zero-point energy and entropy (at 298.15 K), respectively. The zero-point energy and entropy contributions of gas-phase molecules were taken from the NIST database[7] (Table S1), whereas those for adsorbed intermediates were obtained from vibrational frequency calculations processed using VASPKIT.[8] $-eU_{RHE}$ denotes the free energy contribution of electron under potential $U$ referenced to the reversible hydrogen electrode (RHE). In addition, solvation effects were considered using the implicit solvation model implemented in VASPsol.[9]

The OER pathway can be described as follows:[10]

$$* + H_2O \rightarrow OH* + (H^+ + e^-) \quad \text{(a)}$$

$$OH* \rightarrow O* + (H^+ + e^-) \quad \text{(b)}$$

$$H_2O + O* \rightarrow OOH* + (H^+ + e^-) \quad \text{(c)}$$

$$OOH* \rightarrow O_2 + (H^+ + e^-) \quad \text{(d)}$$

The reaction free energy of above four elementary steps can be calculated by

$$\Delta G_a = \Delta G_{*OH} \quad (1)$$

$$\Delta G_b = \Delta G_{*O} - \Delta G_{*OH} \quad (2)$$

$$\Delta G_c = \Delta G_{*OOH} - \Delta G_{*O} \quad (3)$$

$$\Delta G_d = 4.92 - \Delta G_{*OOH} \quad (4)$$

Here, $\Delta G_{*OH}$, $\Delta G_{*O}$, and $\Delta G_{*OOH}$ denote the adsorption free energies of *OH, *O, and *OOH, respectively, and are calculated by referring to the free energy of $H_2O$ and $H_2$,

$$\Delta G_{*OH} = G_{*OH} - (G_{H2O} - 1/2G_{H2}) - G_{*} \quad (5)$$

$$\Delta G_{*O} = G_{*O} - (G_{H2O} - G_{H2}) - G_{*} \quad (6)$$

$$\Delta G_{*OOH} = G_{*OOH} - (2G_{H2O} - 3/2G_{H2}) - G_{*} \quad (7)$$

The theoretical overpotential of OER ($\eta_{OER}$), which is used as an indicator of catalytic activity, is defined as:

$$\eta_{OER} = -1.23 + \max(\Delta G_a, \Delta G_b, \Delta G_c, \Delta G_d)/e \quad (8)$$

The electrode potential effect was treated using the constant-potential (CP) method developed by Duan and Xiao, with potentials referenced to the standard hydrogen electrode (SHE).[11,12] Structural optimization under CP conditions was performed within a grand canonical framework, where atomic configurations are optimized at a fixed electrode potential rather than a fixed electron number. During geometry optimization, the total number of electrons is iteratively adjusted via an outer self-consistent loop to align the Fermi level with the target potential.

The electrode potential of the charged slab relative to the standard hydrogen electrode ($U_{SHE}$) was determined as:

$$U_{SHE} = -4.6 - \phi/\mathrm{e} \quad (9)$$

Here, −4.6 eV corresponds to the work function of the SHE in VASPsol,[13,14] and $-\phi$ denotes the work function of the charged system. In VASPsol, the Debye screening length was set to 3.0 Å and the relative dielectric constant to 80 to mimic the aqueous environment, while the nonelectrostatic parameter TAU was set to zero to avoid numerical instabilities.[15,16]

The relationship between the electrode potential referenced to the RHE and that referenced to the SHE is given by $eU_{RHE} = eU_{SHE} + k_BT \times \mathrm{pH} \times \ln 10$ (pH = 1).[17]

The hydrogen evolution reaction (HER) activity was evaluated based on the hydrogen adsorption free energy ($\Delta G_{*H}$),[18,19] calculated by:

$$\Delta G_{*H} = \Delta E_{*H} + \Delta E_{ZPE} - T\Delta S_H + eU_{RHE} \quad (10)$$

A smaller $|\Delta G_{*H}|$ corresponds to better HER performance, with the optimal activity achieved when $\Delta G_{*H}$ approaches 0 eV.[20]

The exchange current density ($i_0$) was calculated based on Nørskov's assumption, which describes the intrinsic rate of proton transfer from the solvent to the catalytic surface. $i_0$ is expressed as follows:[19,21]

$$i_0 = \frac{-ek_0}{1+\exp(|\Delta G_{*H}/k_b T|)} \quad (11)$$

Here, $k_0$ denotes the reaction rate constant at zero overpotential (set to 1), and $k_b$ is the Boltzmann constant.

The PCET kinetic barrier of the potential determining step (PDS) for the OER was calculated under constant-potential conditions using the electrochemical nudged elastic band (eNEB) method.[22,23] To simulate acidic reaction conditions, three $H_2O$ molecules and one $H_3O^+$ were introduced into the model. The minimum energy pathway (MEP) was determined by interpolating six intermediate images between the fully relaxed initial (IS) and final (FS) states, yielding the transition state (TS) geometry and the kinetic barriers.

Machine learning calculations were performed using a Bayesian optimization-random forest (BO-RF) model implemented with the Scikit-learn library, in which bayesian optimization was employed to efficiently search the hyperparameter space and determine the optimal combination of hyperparameters.[24-26] The dataset was randomly divided into a training set (80%) and a test set (20%). A 5-fold cross-validation was further employed to improve the accuracy of the model.

The predictive performance of the model was evaluated using the coefficient of determination ($R^2$), mean absolute error (MAE), and root mean square error (RMSE), which are defined as follows:

$$R^2 = 1 - \frac{\sum_{i=1}^{n}(y_i - \hat{y}_i)^2}{\sum_{i=1}^{n}(y_i - \bar{y})^2} \quad (12)$$

$$MAE = \frac{1}{n}\sum_{i=1}^{n}|y_i - \hat{y}_i| \quad (13)$$

$$RMSE = \sqrt{\frac{1}{n}\sum_{i=1}^{n}(y_i - \hat{y}_i)^2} \quad (14)$$

where $y_i$ and $\hat{y}_i$ represent the true and predicted values of the $i$-th sample, respectively, $\bar{y}$ is the mean of the true values, and $n$ is the total number of samples. The closer the values of RMSE and MAE are to 0, and the closer the value of $R^2$ is to 1, the better the predictive performance of the machine learning model.

The Sure Independence Screening and Sparsifying Operator (SISSO) algorithm was employed to determine effective descriptors for predicting $\eta_{OER}$ of $TM_1TM_2$-$C_{6-x}N_x$ DACs. SISSO is a machine learning method that integrates symbolic regression with compressed sensing. High-dimensional feature spaces are first reduced via Sure Independence Screening (SIS), which evaluates the correlations between features and the target property to select highly relevant feature subsets. The optimal n-dimensional descriptors are then determined through Sequential Optimization (SO). SISSO offers advantages in its simplicity, physical interpretability, and robustness on small datasets, without requiring additional feature-selection procedures or large amounts of training data. In contrast to complex black-box models that yield abstract feature-importance metrics, SISSO directly produces low-dimensional mathematical expressions that are physically meaningful and readily interpretable within the framework of materials science.[27]

The feasibility of experimental synthesis is evaluated by calculating the formation energy ($E_f$) according to the following equation:

$$E_f = E_{total} - E_G + \sum n_i\mu_i \quad (15)$$

where, $E_{total}$ and $E_G$ are the total energies of the $TM_1TM_2$-$C_{6-x}N_x$ DACs and the 6×3 graphene supercells, respectively. $\mu_i$ represents the chemical potential of the $i$ species. $n_i$ denotes the number of the removed or added $i$ species, where the positive and negative values are referred to the removed and added species, respectively. The chemical potentials of the involved species, $\mu_C$, $\mu_N$, and $\mu_{TM}$, are calculated using pristine graphene unit cell, $N_2$ free molecule, and metal bulk, respectively.

The thermodynamic stability is evaluated by calculating the binding energy ($E_b$) as follows:

$$E_b = E_{total} - E_{CN} - E_{TM1} - E_{TM2} \quad (16)$$

Here, $E_{CN}$ and $E_{total}$ are the total energies of the N-doped graphene and the $TM_1TM_2$-$C_{6-x}N_x$ DACs, respectively. $E_{TM1}$ and $E_{TM2}$ are the total energies of the isolated TM atoms. According to this definition, a more negative $E_b$ indicates stronger binding between $E_{TM1}/E_{TM2}$ and the graphene substrate, and thus greater structural stability.

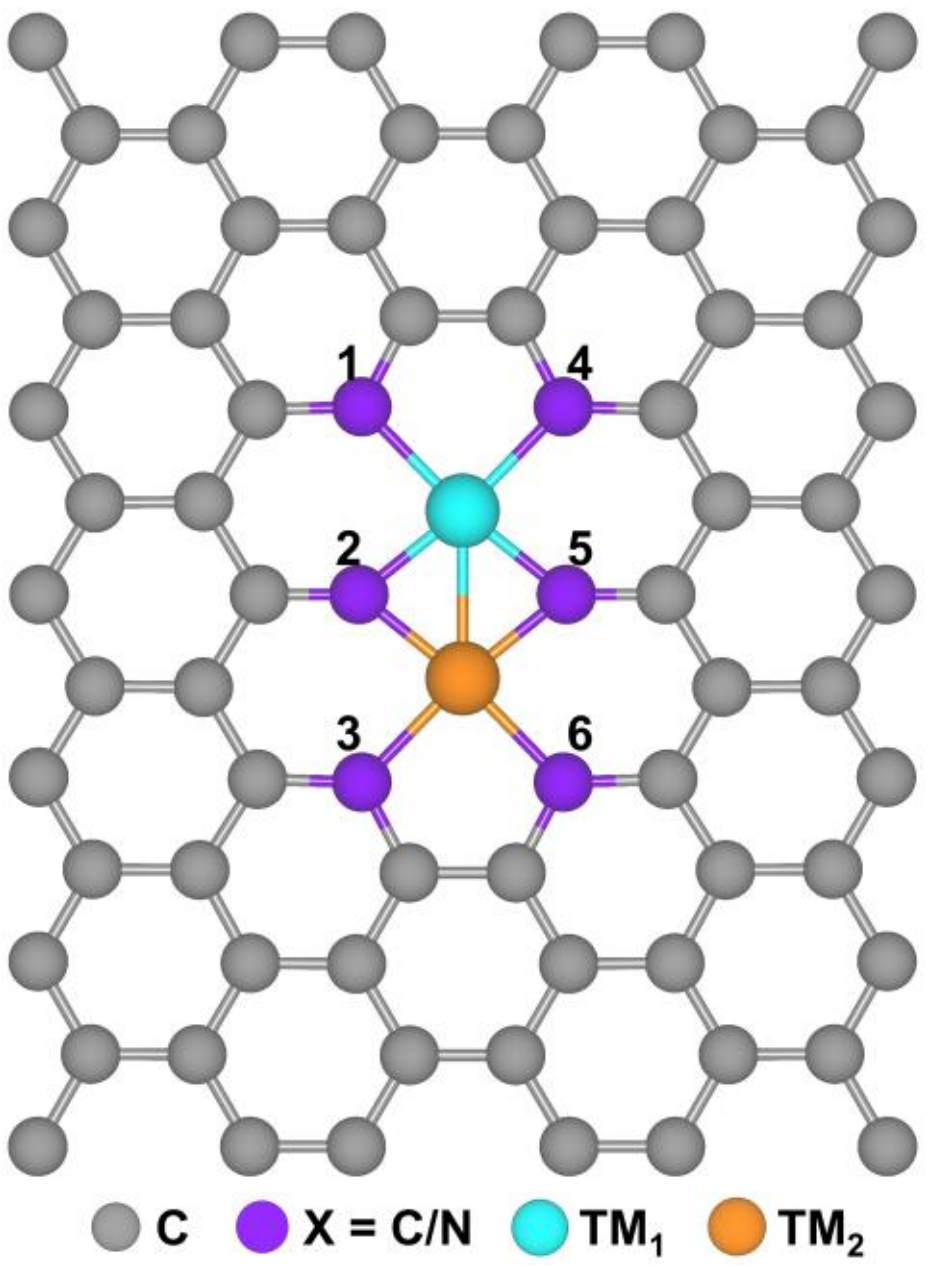

**Fig. S1.** Schematic of $TM_1TM_2$-$C_{6-x}N_x$ DACs, showing the coding of first-shell coordination atoms. Gray, purple, cyan, and brown spheres represent C, X (X = C, N), $TM_1$, and $TM_2$, respectively.

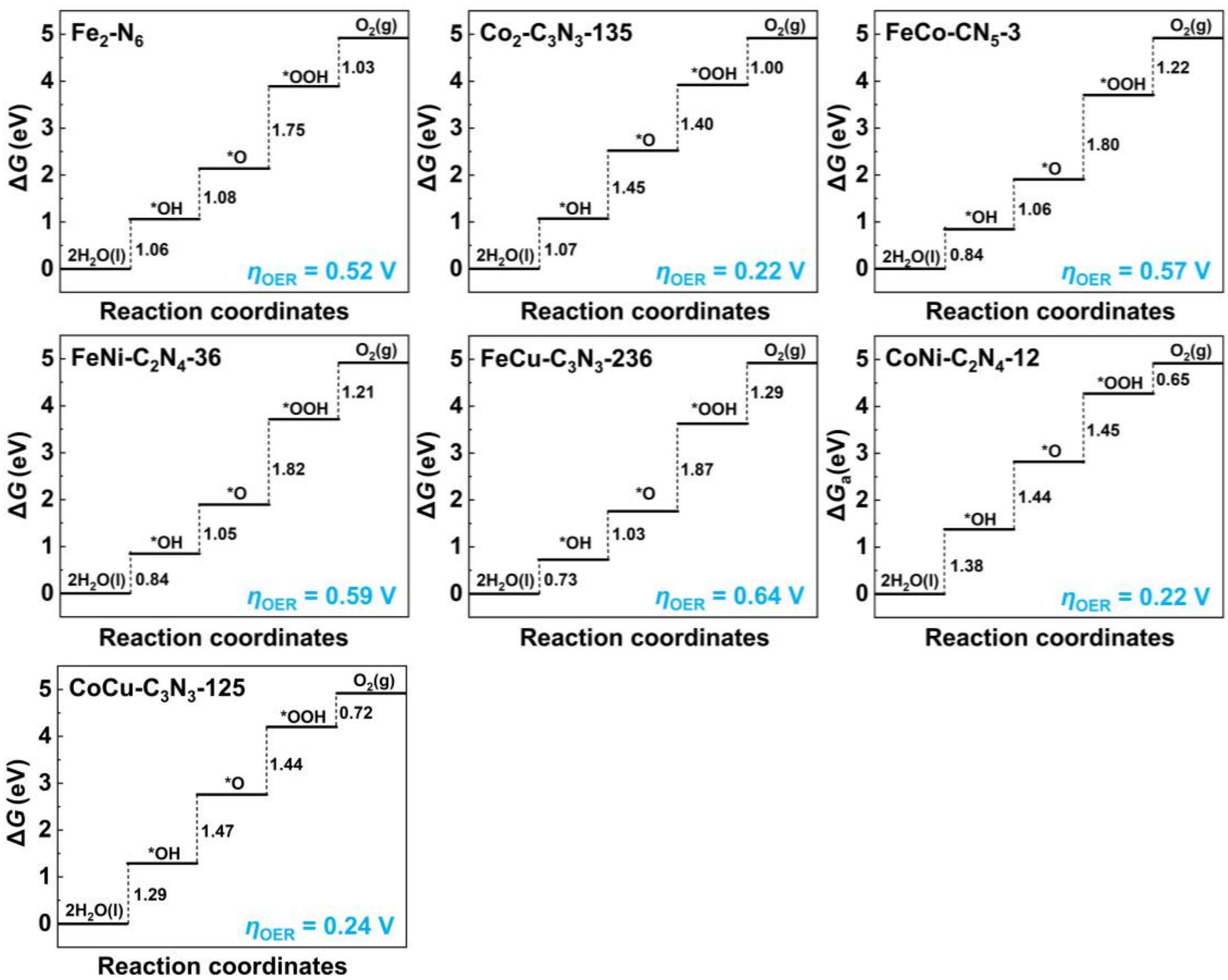

**Fig. S2.** OER free energy diagrams for the most active coordination configurations in seven $TM_1TM_2$-$C_{6-x}N_x$ DACs systems ($TM_1TM_2$ = $Fe_2$, $Co_2$, FeCo, FeNi, FeCu, CoNi, and CoCu).

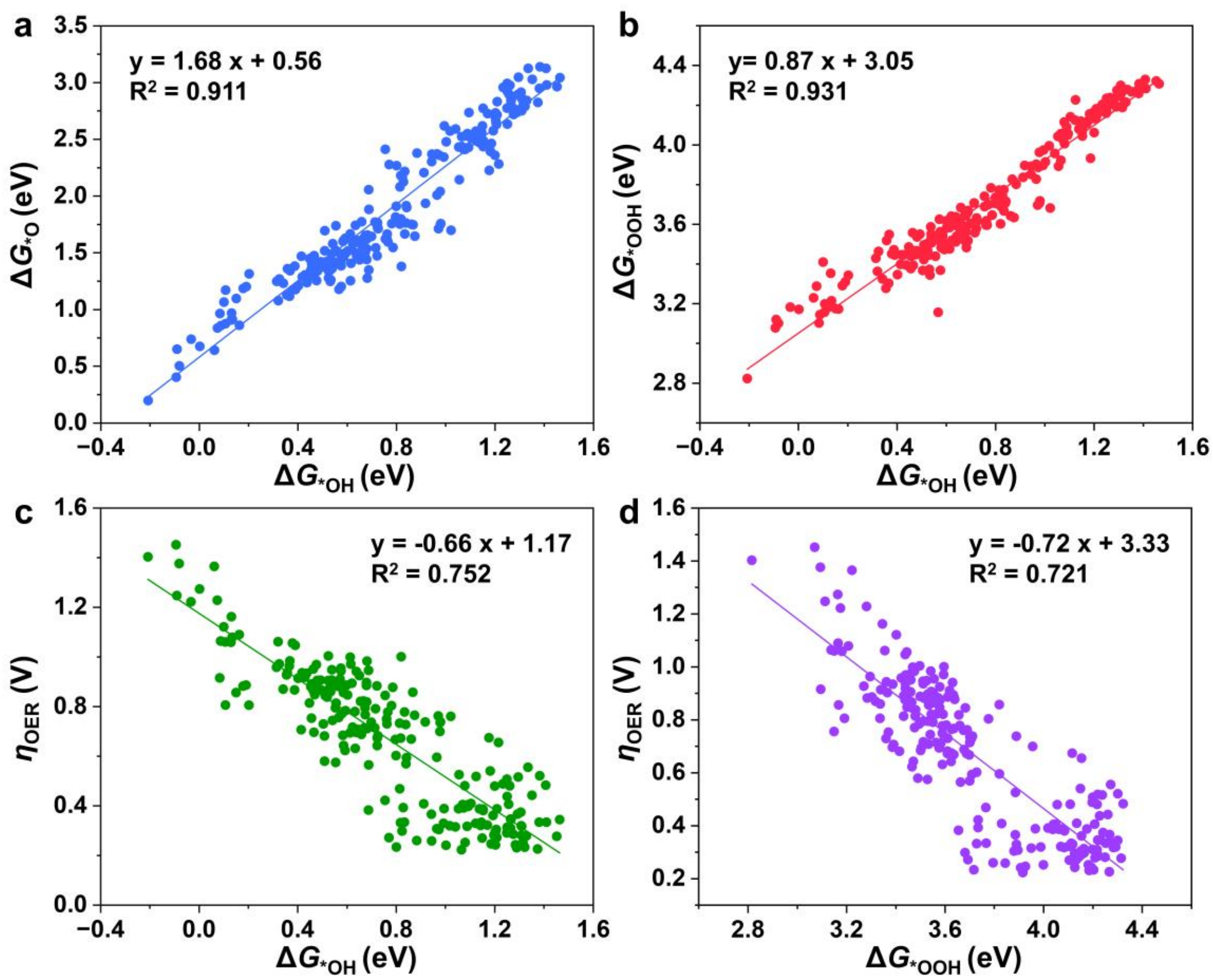

**Fig. S3.** Linear relationships of $\Delta G_{*O}$ (a) and $\Delta G_{*OOH}$ (b) versus $\Delta G_{*OH}$, and correlations of $\Delta G_{*OH}$ (c) and $\Delta G_{*OOH}$ (d) with the OER overpotential ($\eta_{OER}$) for all $TM_1TM_2$-$C_{6-x}N_x$ DACs.

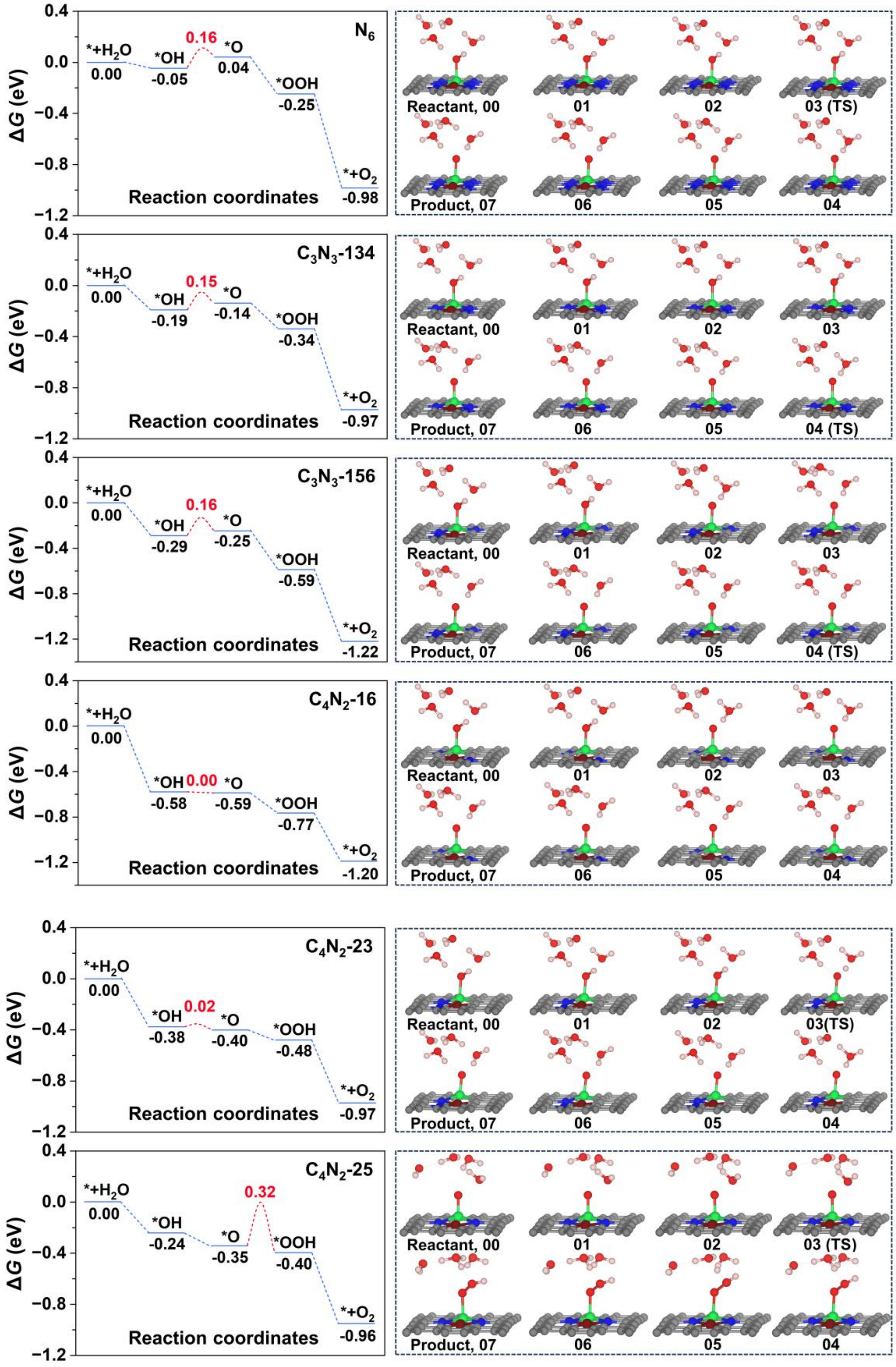

ΔG (eV)
Reaction coordinates
N6
*+H2O
0.00
*OH
-0.05
0.16
*O
0.04
*OOH
-0.25
*+O2
-0.98
Reactant, 00
01
02
03 (TS)
Product, 07
06
05
04
C3N3-134
0.00
-0.19
0.15
-0.14
-0.34
-0.97
04 (TS)
C3N3-156
0.00
-0.29
0.16
-0.25
-0.59
-1.22
C4N2-16
0.00
-0.58
0.00
-0.59
-0.77
-1.20
C4N2-23
0.00
-0.38
0.02
-0.40
-0.48
-0.97
03(TS)
C4N2-25
0.00
-0.24
-0.35
0.32
-0.40
-0.96

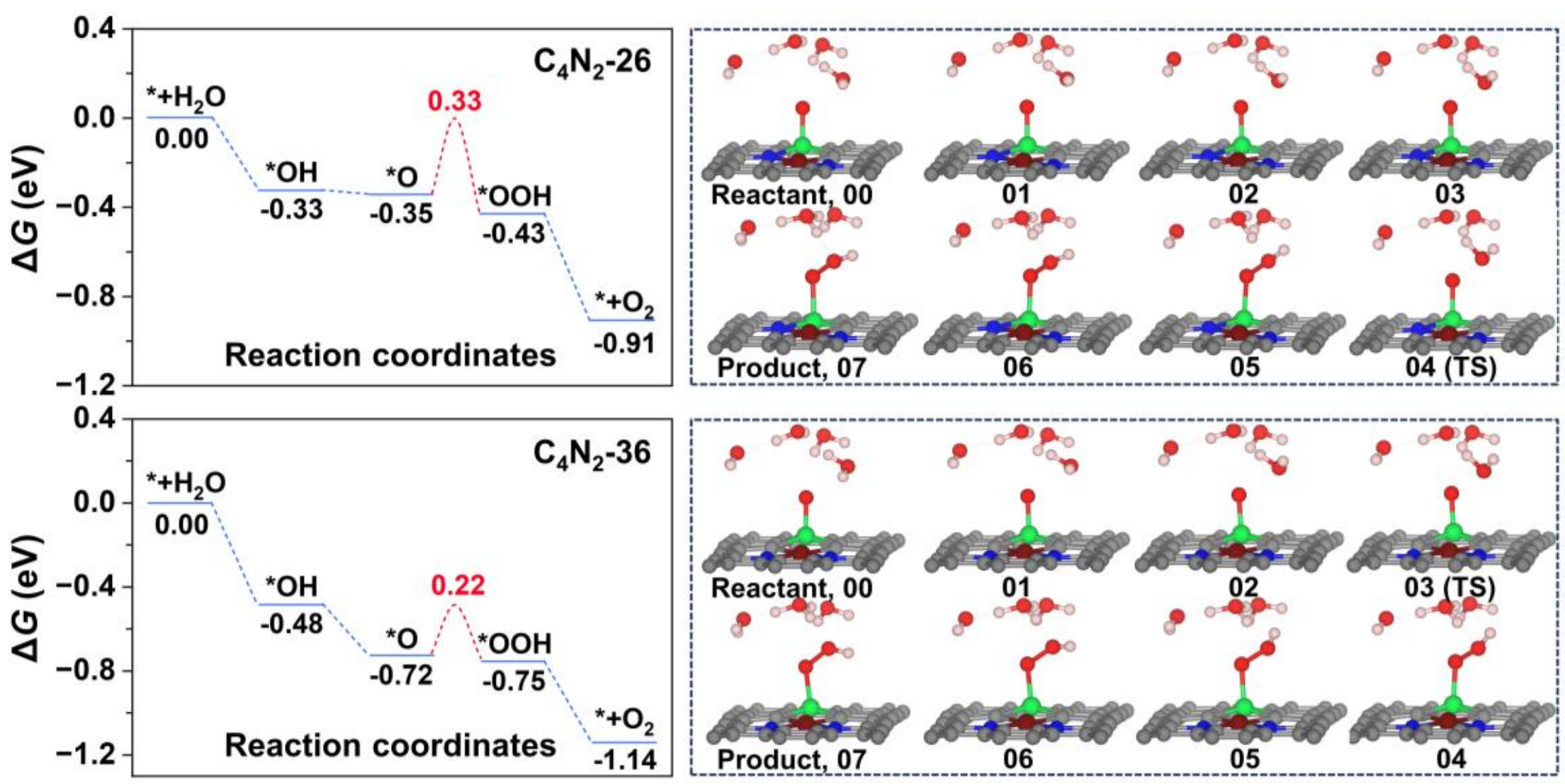

**Fig. S4.** OER free energy diagrams (left) and PCET kinetic barriers of the potential-determining steps (PDSs) (right) for CoNi-$C_{6-x}N_x$, evaluated at $U_{RHE} = \Delta G_{max}/e$.

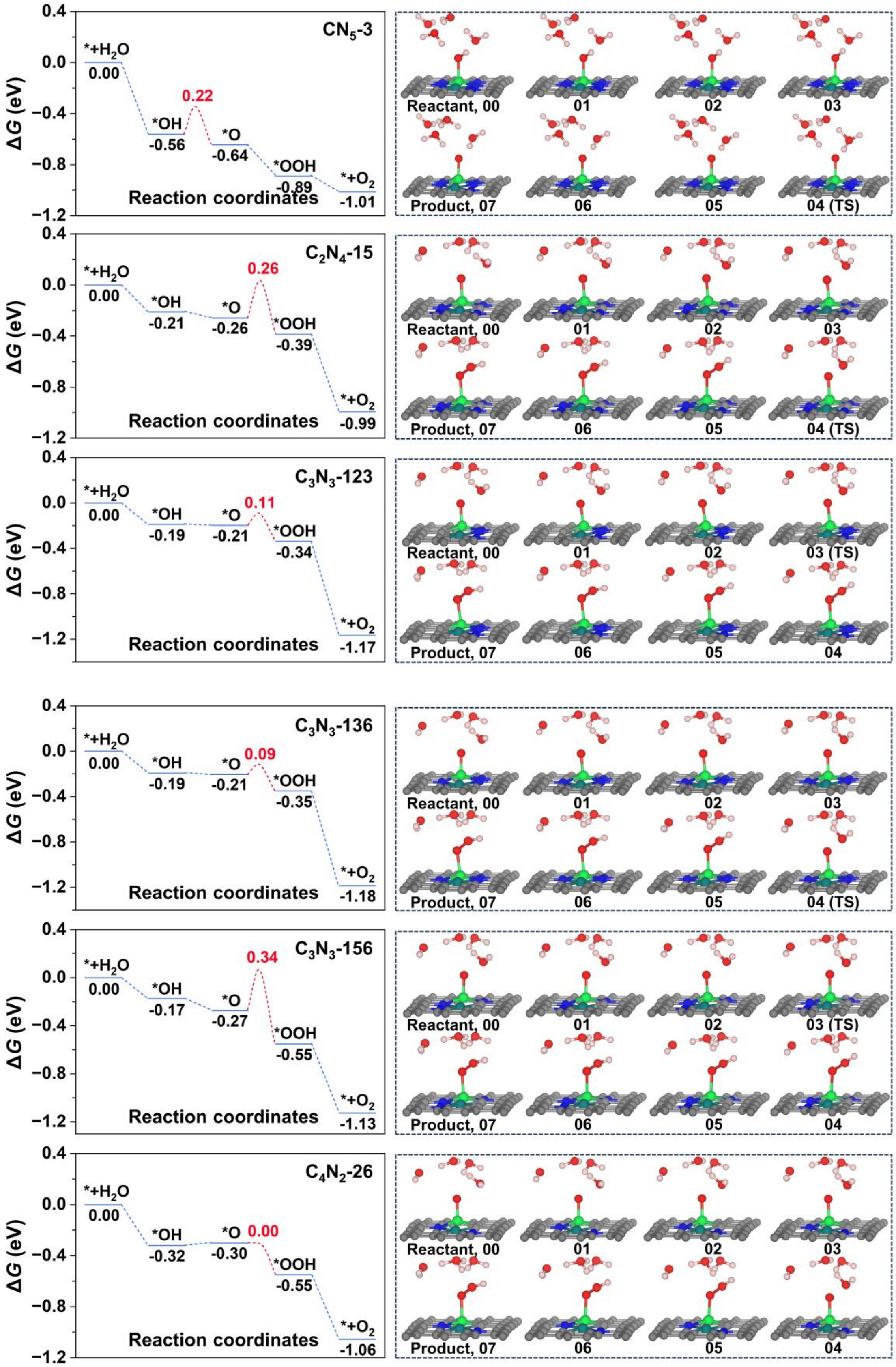

CN$_5$-3
*+H$_2$O
0.00
*OH
-0.56
0.22
*O
-0.64
*OOH
-0.89
*+O$_2$
-1.01
Reaction coordinates
ΔG (eV)
Reactant, 00
01
02
03
Product, 07
06
05
04 (TS)
C$_2$N$_4$-15
*+H$_2$O
0.00
*OH
-0.21
*O
-0.26
0.26
*OOH
-0.39
*+O$_2$
-0.99
C$_3$N$_3$-123
*+H$_2$O
0.00
*OH
-0.19
*O
-0.21
0.11
*OOH
-0.34
*+O$_2$
-1.17
03 (TS)
04
C$_3$N$_3$-136
*+H$_2$O
0.00
*OH
-0.19
*O
-0.21
0.09
*OOH
-0.35
*+O$_2$
-1.18
C$_3$N$_3$-156
*+H$_2$O
0.00
*OH
-0.17
*O
-0.27
0.34
*OOH
-0.55
*+O$_2$
-1.13
C$_4$N$_2$-26
*+H$_2$O
0.00
*OH
-0.32
*O
-0.30
0.00
*OOH
-0.55
*+O$_2$
-1.06

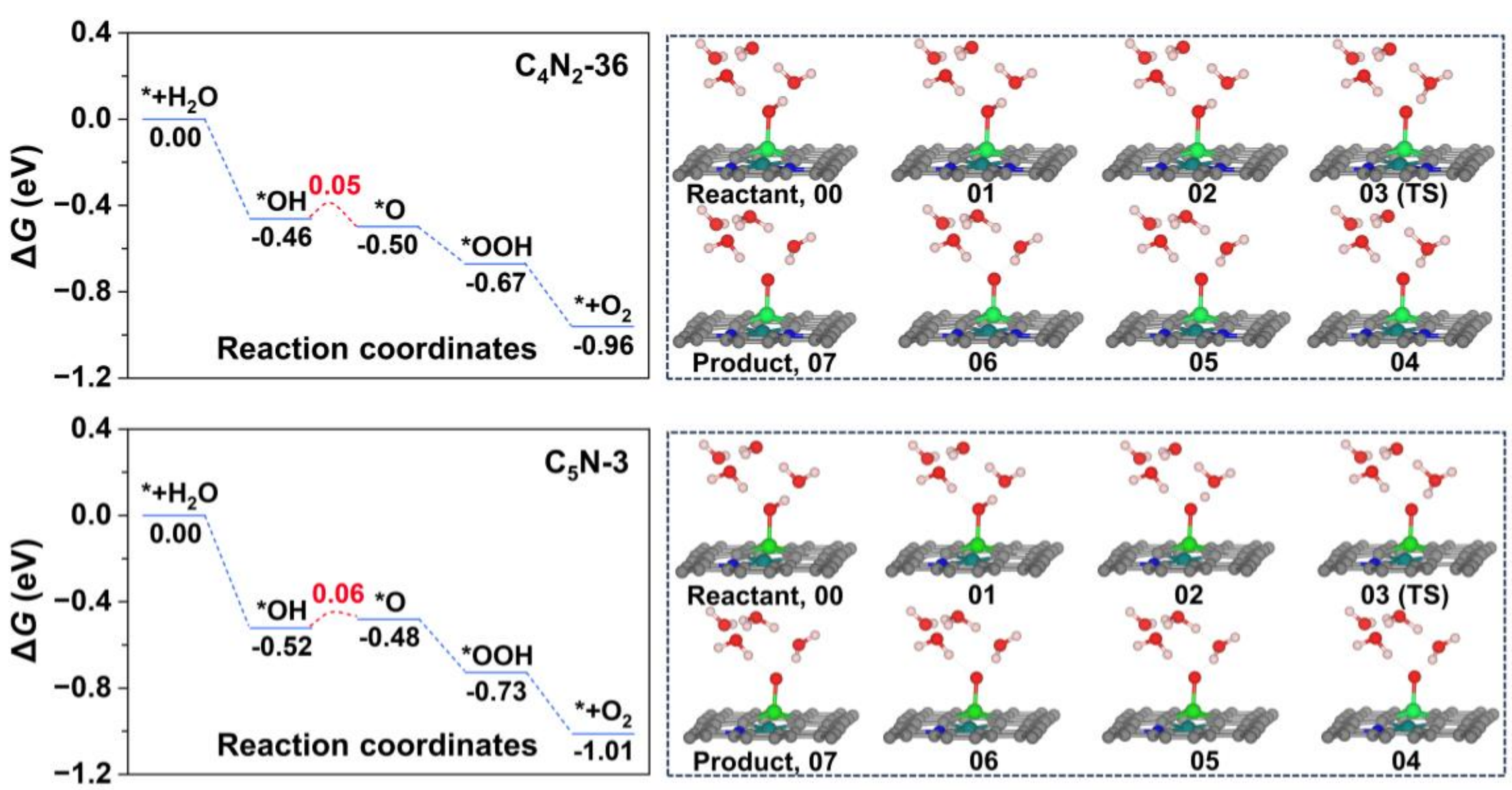

**Fig. S5.** OER free energy diagrams (left) and PCET kinetic barriers of the PDSs (right) for CoCu-$C_{6-x}N_x$ DACs, evaluated at $U_{RHE} = \Delta G_{max}/e$.

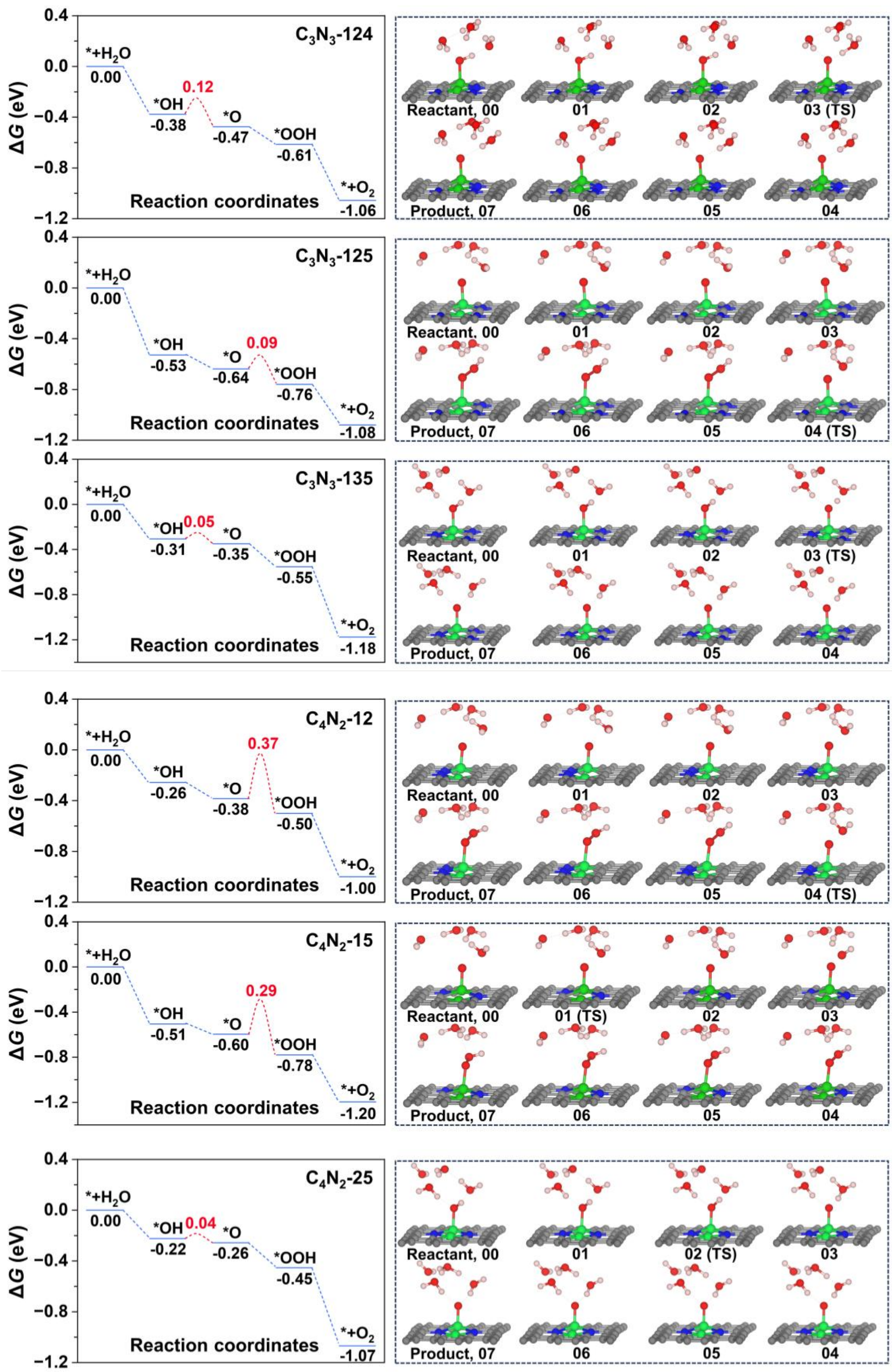

**Fig. S6.** OER free energy diagrams (left) and PCET kinetic barriers of the PDSs (right) for $Co_2$-$C_{6-x}N_x$ DACs, evaluated at $U_{RHE} = \Delta G_{max}/e$.

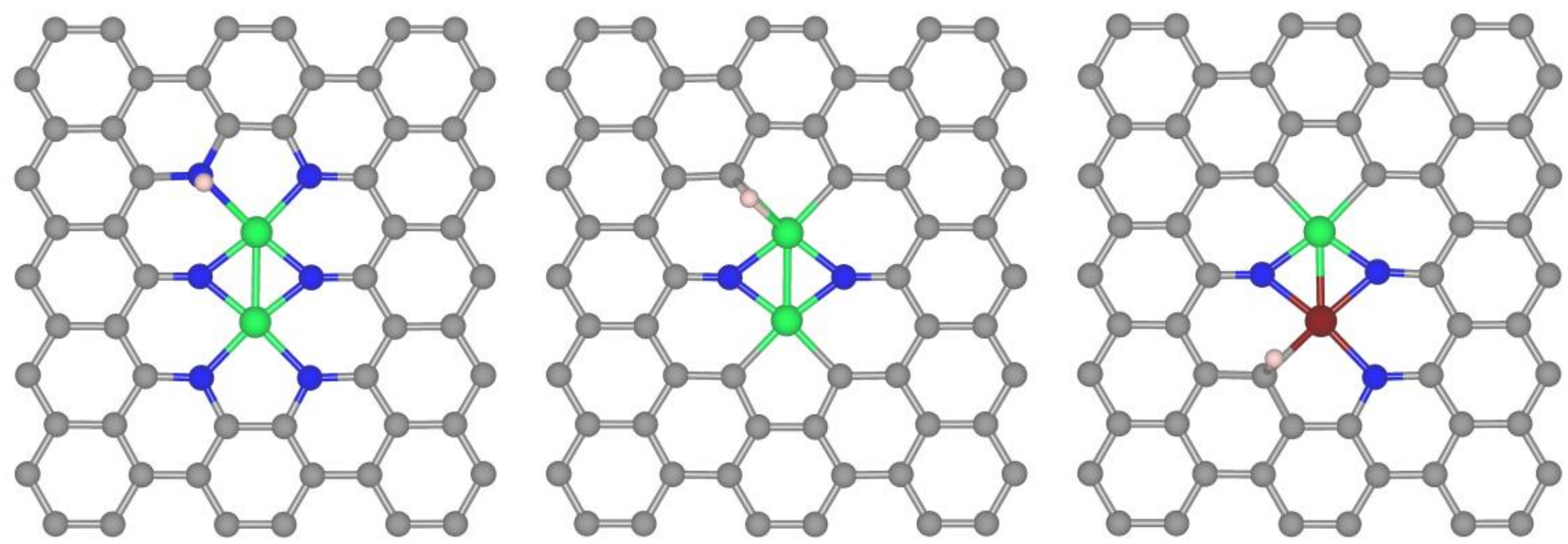

**Fig. S7.** Optimized structures of H adsorption at the N top site on $Co_2$-$N_6$ (left), the Co–C bridge site on $Co_2$–$C_4N_2$-25 (middle), and the C top site on CoNi–$C_3N_3$-134 (right).

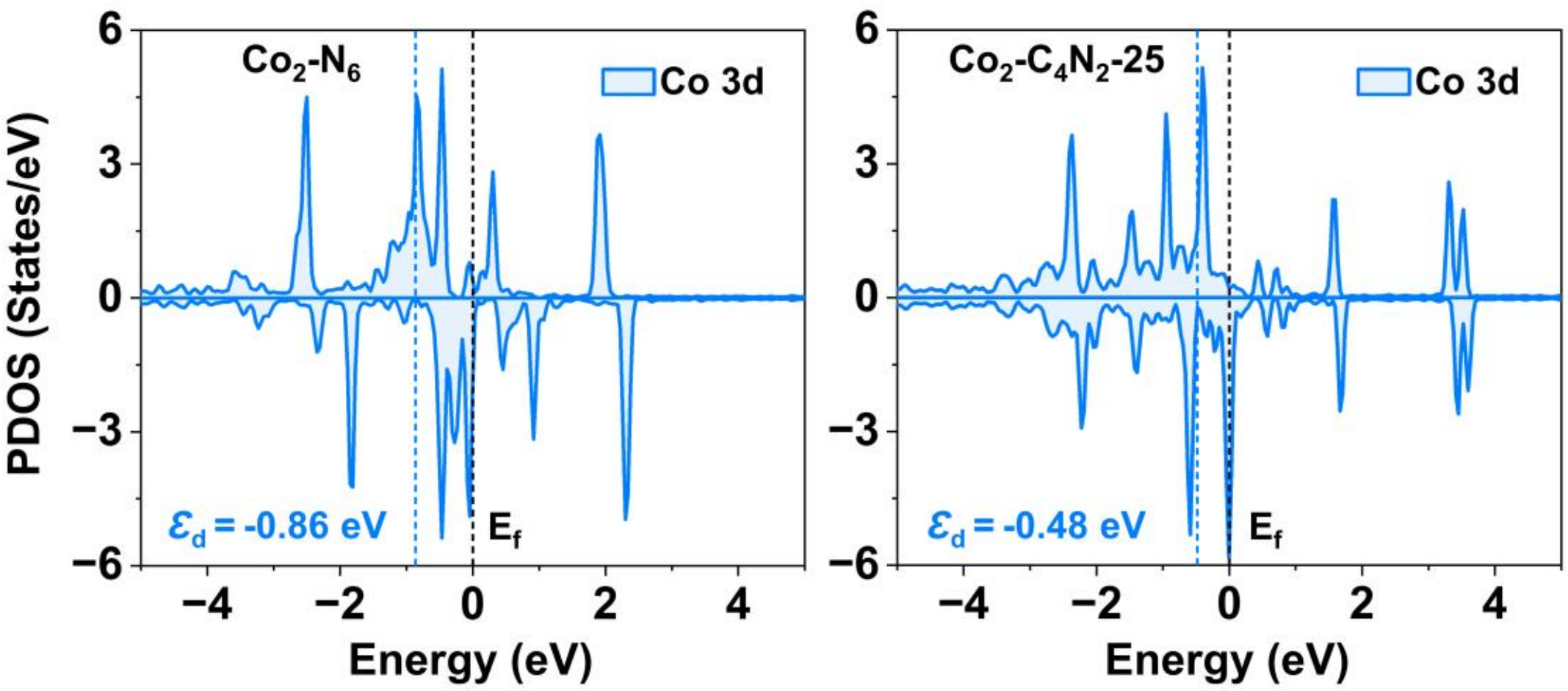

**Fig. S8.** Projected density of states (PDOS) of the Co 3*d* orbitals for $Co_2$-$N_6$ (left) and $Co_2$-$C_4N_2$-25 (right). Positive and negative PDOS values represent the spin-up and spin-down states, respectively, and the vertical lines indicate the positions of the Fermi level ($E_f$) and the d-band center ($\varepsilon_d$).

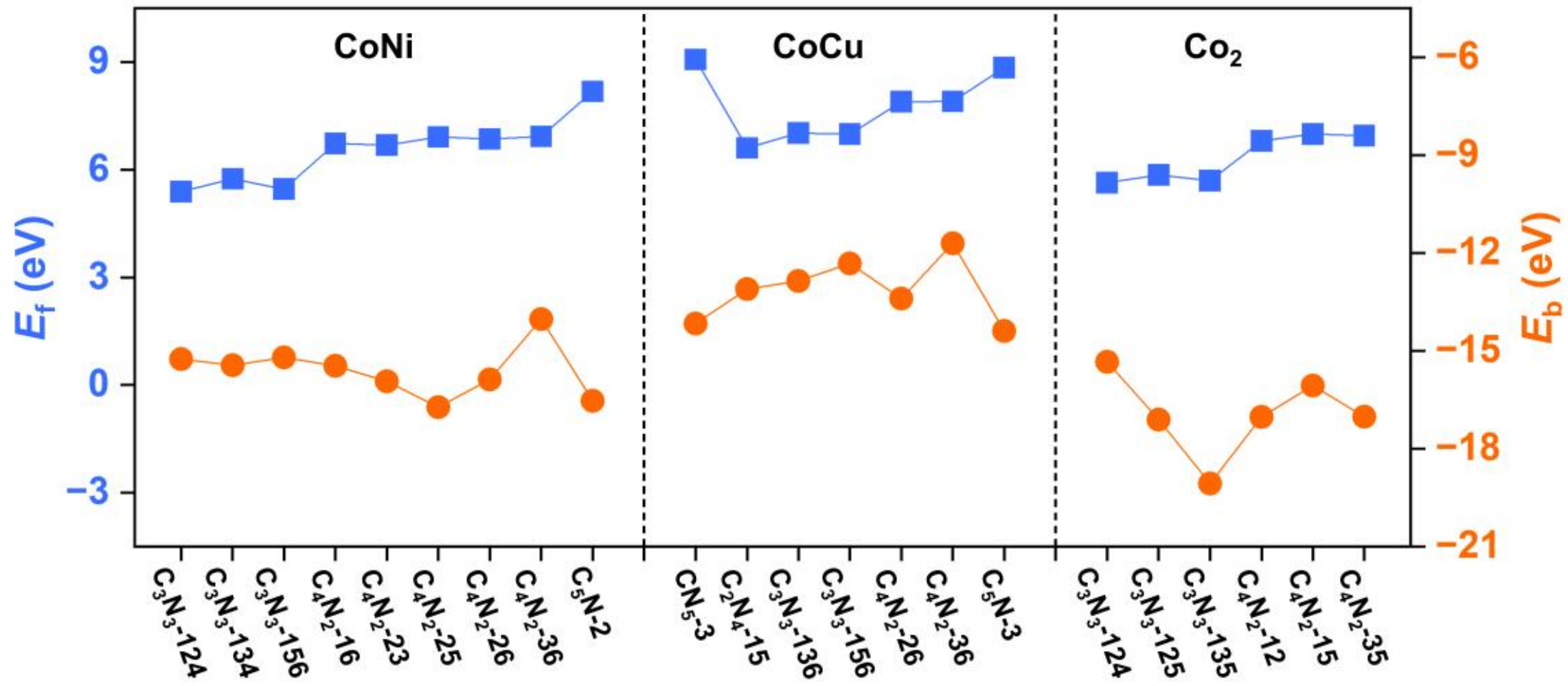

**Fig. S9.** Calculated formation ($E_f$) and binding energies ($E_b$) of candidate $TM_1TM_2$-$C_{6-x}N_x$ DACs.

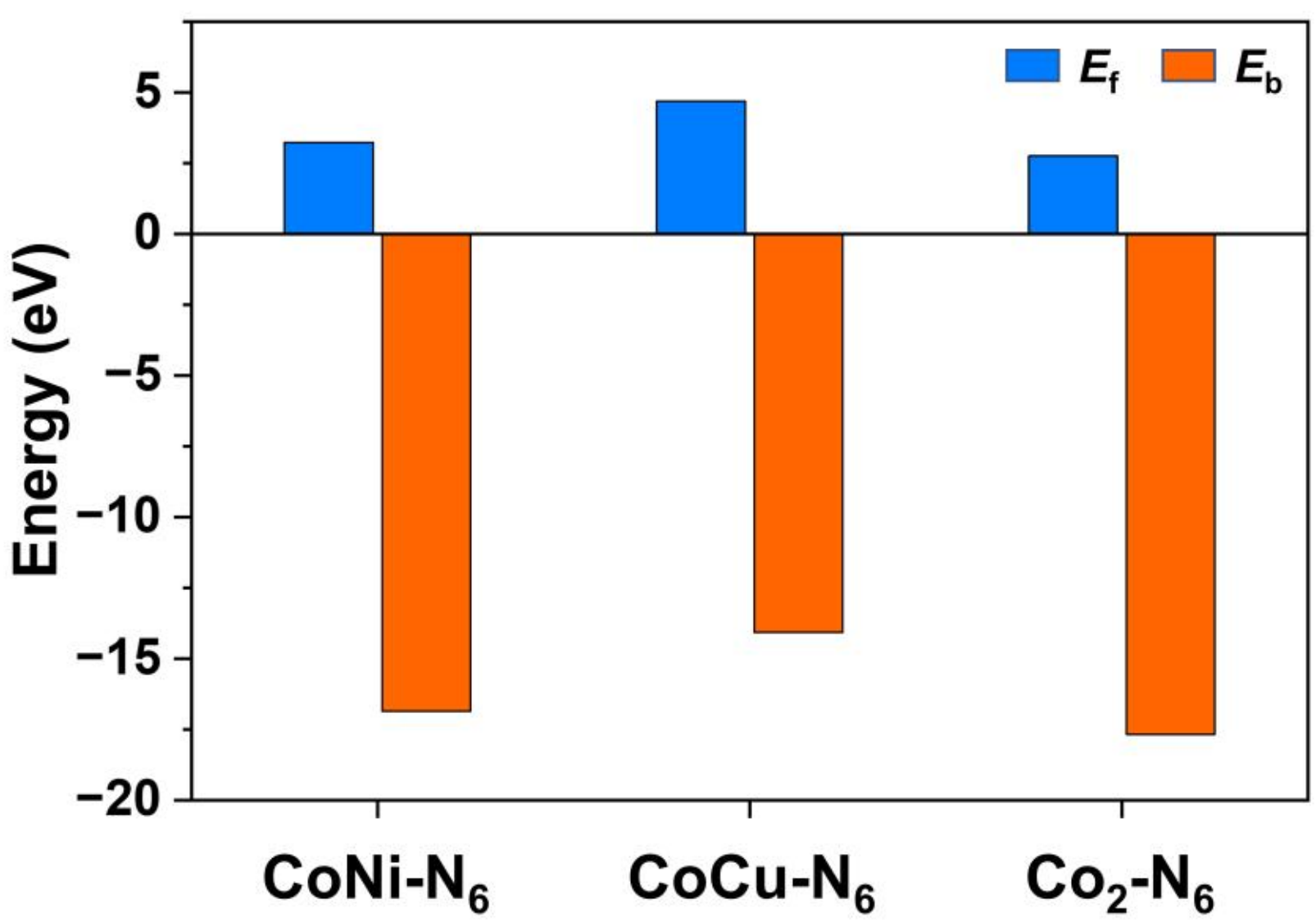

**Fig. S10.** Calculated formation energy ($E_f$) and binding energy ($E_b$) of $CoTM_2$-$N_6$ DACs.

**Table S1.** Zero-point energy ($E_{ZPE}$, eV), entropic contribution ($TS$, $T$ = 298.15 K, eV), and solvation correction ($\Delta E_{sol}$) for OER and HER intermediates on $TM_1TM_2$-$C_{6-x}N_x$ DACs; * represents the adsorption site.

| Species | $E_{ZPE}$ | $TS$ | $\Delta E_{sol}$ |
|---|---|---|---|
| $H_2O$ | 0.56 | 0.67 | -- |
| $H_2$ | 0.27 | 0.40 | -- |
| *OH | 0.34 | 0.07 | -0.22 |
| *O | 0.07 | 0.06 | -0.29 |
| *OOH | 0.44 | 0.13 | -0.26 |

**Table S2.** Calculated adsorption free energies of *OH, *O, and *OOH on $Fe_2$-$C_{6-x}N_x$ DACs.

| | $\Delta G_{*OH}$ (eV) | $\Delta G_{*O}$ (eV) | $\Delta G_{*OOH}$ (eV) |
|---|---|---|---|
| $N_6$ | 1.06 | 2.14 | 3.89 |
| $CN_5$-1 | 0.87 | 1.74 | 3.82 |
| $CN_5$-2 | 0.79 | 1.75 | 3.78 |
| $C_2N_4$-12 | 0.64 | 1.47 | 3.63 |
| $C_2N_4$-13 | 0.79 | 1.54 | 3.64 |
| $C_2N_4$-14 | 0.68 | 1.49 | 3.52 |
| $C_2N_4$-15 | 0.57 | 1.41 | 3.52 |
| $C_2N_4$-16 | 0.68 | 1.53 | 3.64 |
| $C_2N_4$-25 | 0.50 | 1.37 | 3.46 |
| $C_3N_3$-123 | 0.62 | 1.25 | 3.47 |
| $C_3N_3$-124 | 0.54 | 1.25 | 3.34 |
| $C_3N_3$-125 | 0.57 | 1.17 | 3.15 |
| $C_3N_3$-126 | 0.69 | 1.27 | 3.48 |
| $C_3N_3$-134 | 0.58 | 1.38 | 3.56 |
| $C_3N_3$-135 | 0.58 | 1.19 | 3.36 |
| $C_4N_2$-12 | 0.54 | 1.24 | 3.37 |
| $C_4N_2$-13 | 0.11 | 0.86 | 3.15 |
| $C_4N_2$-14 | 0.09 | 0.96 | 3.10 |
| $C_4N_2$-15 | 0.41 | 1.20 | 3.34 |
| $C_4N_2$-16 | 0.09 | 0.85 | 3.14 |
| $C_4N_2$-25 | 0.53 | 1.33 | 3.46 |
| $C_5N$-1 | -0.09 | 0.64 | 3.12 |
| $C_5N$-2 | 0.17 | 0.85 | 3.17 |
| $C_6$ | -0.21 | 0.19 | 2.82 |

**Table S3.** Calculated adsorption free energies of *OH, *O, and *OOH on $Co_2$-$C_{6-x}N_x$ DACs.

| | $\Delta G_{*OH}$ (eV) | $\Delta G_{*O}$ (eV) | $\Delta G_{*OOH}$ (eV) |
|---|---|---|---|
| $N_6$ | 1.36 | 3.02 | 4.27 |
| $CN_5$-1 | 1.46 | 2.96 | 4.32 |
| $CN_5$-2 | 1.41 | 3.12 | 4.33 |
| $C_2N_4$-12 | 1.30 | 2.89 | 4.27 |
| $C_2N_4$-13 | 1.42 | 2.97 | 4.28 |
| $C_2N_4$-14 | 1.27 | 2.81 | 4.20 |
| $C_2N_4$-15 | 1.32 | 2.80 | 4.19 |
| $C_2N_4$-16 | 1.39 | 2.95 | 4.26 |
| $C_2N_4$-25 | 1.10 | 2.73 | 4.05 |
| $C_3N_3$-123 | 1.00 | 2.61 | 3.97 |
| $C_3N_3$-124 | 0.95 | 2.36 | 3.85 |
| $C_3N_3$-125 | 0.83 | 2.16 | 3.68 |
| $C_3N_3$-126 | 1.02 | 2.57 | 3.99 |
| $C_3N_3$-134 | 1.08 | 2.53 | 4.00 |
| $C_3N_3$-135 | 1.07 | 2.52 | 3.92 |
| $C_4N_2$-12 | 0.95 | 2.30 | 3.89 |
| $C_4N_2$-13 | 0.51 | 1.69 | 3.49 |
| $C_4N_2$-14 | 0.69 | 2.05 | 3.66 |
| $C_4N_2$-15 | 0.92 | 2.20 | 3.83 |
| $C_4N_2$-16 | 0.56 | 1.73 | 3.53 |
| $C_4N_2$-25 | 1.00 | 2.34 | 3.89 |
| $C_5N$-1 | 0.60 | 1.62 | 3.47 |
| $C_5N$-2 | 0.64 | 1.73 | 3.59 |
| $C_6$ | 0.47 | 1.47 | 3.39 |

**Table S4.** Calculated adsorption free energies of *OH, *O, and *OOH on FeCo-$C_{6-x}N_x$ DACs.

| | $\Delta G_{*OH}$ (eV) | $\Delta G_{*O}$ (eV) | $\Delta G_{*OOH}$ (eV) |
|---|---|---|---|
| $N_6$ | 0.97 | 2.00 | 3.82 |
| $CN_5$-1 | 0.98 | 1.75 | 3.71 |
| $CN_5$-2 | 0.82 | 1.79 | 3.69 |
| $CN_5$-3 | 0.84 | 1.90 | 3.70 |
| $C_2N_4$-12 | 0.68 | 1.61 | 3.66 |
| $C_2N_4$-13 | 0.98 | 1.74 | 3.70 |
| $C_2N_4$-14 | 0.77 | 1.59 | 3.59 |
| $C_2N_4$-15 | 0.67 | 1.55 | 3.60 |
| $C_2N_4$-16 | 0.98 | 1.70 | 3.69 |
| $C_2N_4$-23 | 0.84 | 1.76 | 3.71 |
| $C_2N_4$-25 | 0.60 | 1.50 | 3.61 |
| $C_2N_4$-26 | 0.72 | 1.76 | 3.70 |
| $C_2N_4$-36 | 0.80 | 1.90 | 3.73 |
| $C_3N_3$-123 | 0.82 | 1.37 | 3.60 |
| $C_3N_3$-124 | 0.47 | 1.29 | 3.46 |
| $C_3N_3$-125 | 0.37 | 1.11 | 3.30 |
| $C_3N_3$-126 | 0.62 | 1.40 | 3.60 |
| $C_3N_3$-134 | 0.62 | 1.52 | 3.64 |
| $C_3N_3$-135 | 0.54 | 1.35 | 3.53 |
| $C_3N_3$-136 | 0.76 | 1.61 | 3.69 |
| $C_3N_3$-156 | 0.69 | 1.34 | 3.51 |
| $C_3N_3$-235 | 0.42 | 1.28 | 3.44 |
| $C_3N_3$-236 | 0.54 | 1.41 | 3.55 |
| $C_4N_2$-12 | 0.48 | 1.33 | 3.46 |
| $C_4N_2$-13 | 0.13 | 0.90 | 3.18 |
| $C_4N_2$-14 | 0.36 | 1.12 | 3.27 |
| $C_4N_2$-15 | 0.48 | 1.25 | 3.44 |
| $C_4N_2$-16 | 0.14 | 0.90 | 3.21 |
| $C_4N_2$-23 | 0.39 | 1.17 | 3.44 |
| $C_4N_2$-25 | 0.58 | 1.48 | 3.62 |
| $C_4N_2$-26 | 0.38 | 1.16 | 3.45 |
| $C_4N_2$-36 | 0.00 | 0.67 | 3.17 |
| $C_5N$-1 | -0.03 | 0.73 | 3.18 |
| $C_5N$-2 | 0.32 | 1.07 | 3.36 |
| $C_5N$-3 | -0.08 | 0.49 | 3.10 |
| $C_6$ | -0.09 | 0.39 | 3.07 |

**Table S5.** Calculated adsorption free energies of *OH, *O, and *OOH on FeNi-$C_{6-x}N_x$ DACs.

| | $\Delta G_{*OH}$ (eV) | $\Delta G_{*O}$ (eV) | $\Delta G_{*OOH}$ (eV) |
|---|---|---|---|
| $N_6$ | 0.70 | 1.64 | 3.60 |
| $CN_5$-1 | 0.65 | 1.43 | 3.58 |
| $CN_5$-2 | 0.73 | 1.73 | 3.63 |
| $CN_5$-3 | 0.82 | 1.75 | 3.71 |
| $C_2N_4$-12 | 0.64 | 1.60 | 3.61 |
| $C_2N_4$-13 | 0.68 | 1.65 | 3.64 |
| $C_2N_4$-14 | 0.68 | 1.43 | 3.56 |
| $C_2N_4$-15 | 0.59 | 1.45 | 3.54 |
| $C_2N_4$-16 | 0.88 | 1.64 | 3.63 |
| $C_2N_4$-23 | 0.73 | 1.70 | 3.61 |
| $C_2N_4$-25 | 0.64 | 1.59 | 3.51 |
| $C_2N_4$-26 | 0.73 | 1.71 | 3.61 |
| $C_2N_4$-36 | 0.84 | 1.89 | 3.71 |
| $C_3N_3$-123 | 0.77 | 1.66 | 3.61 |
| $C_3N_3$-124 | 0.56 | 1.39 | 3.43 |
| $C_3N_3$-125 | 0.53 | 1.34 | 3.44 |
| $C_3N_3$-126 | 0.77 | 1.62 | 3.59 |
| $C_3N_3$-134 | 0.66 | 1.50 | 3.55 |
| $C_3N_3$-135 | 0.72 | 1.54 | 3.55 |
| $C_3N_3$-136 | 0.84 | 1.66 | 3.66 |
| $C_3N_3$-156 | 0.67 | 1.52 | 3.57 |
| $C_3N_3$-235 | 0.55 | 1.57 | 3.54 |
| $C_3N_3$-236 | 0.87 | 1.74 | 3.64 |
| $C_4N_2$-12 | 0.47 | 1.46 | 3.55 |
| $C_4N_2$-13 | 0.18 | 1.18 | 3.29 |
| $C_4N_2$-14 | 0.32 | 1.24 | 3.43 |
| $C_4N_2$-15 | 0.48 | 1.41 | 3.53 |
| $C_4N_2$-16 | 0.19 | 1.19 | 3.31 |
| $C_4N_2$-23 | 0.37 | 1.33 | 3.54 |
| $C_4N_2$-25 | 0.61 | 1.51 | 3.58 |
| $C_4N_2$-26 | 0.36 | 1.35 | 3.51 |
| $C_4N_2$-36 | 0.15 | 1.09 | 3.17 |
| $C_5N$-1 | 0.13 | 0.96 | 3.35 |
| $C_5N$-2 | 0.33 | 1.26 | 3.46 |
| $C_5N$-3 | 0.08 | 0.83 | 3.28 |
| $C_6$ | 0.06 | 0.63 | 3.22 |

**Table S6.** Calculated adsorption free energies of *OH, *O, and *OOH on FeCu-$C_{6-x}N_x$ DACs.

| | $\Delta G_{*OH}$ (eV) | $\Delta G_{*O}$ (eV) | $\Delta G_{*OOH}$ (eV) |
|---|---|---|---|
| $N_6$ | 0.50 | 1.41 | 3.55 |
| $CN_5$-1 | 0.42 | 1.46 | 3.39 |
| $CN_5$-2 | 0.51 | 1.51 | 3.42 |
| $CN_5$-3 | 1.03 | 1.69 | 3.68 |
| $C_2N_4$-12 | 0.54 | 1.48 | 3.58 |
| $C_2N_4$-13 | 0.73 | 1.46 | 3.58 |
| $C_2N_4$-14 | 0.58 | 1.41 | 3.52 |
| $C_2N_4$-15 | 0.43 | 1.30 | 3.44 |
| $C_2N_4$-16 | 0.67 | 1.63 | 3.57 |
| $C_2N_4$-23 | 0.60 | 1.49 | 3.50 |
| $C_2N_4$-25 | 0.65 | 1.60 | 3.54 |
| $C_2N_4$-26 | 0.80 | 1.81 | 3.70 |
| $C_2N_4$-36 | 0.69 | 1.65 | 3.59 |
| $C_3N_3$-123 | 0.45 | 1.44 | 3.48 |
| $C_3N_3$-124 | 0.44 | 1.23 | 3.39 |
| $C_3N_3$-125 | 0.52 | 1.40 | 3.39 |
| $C_3N_3$-126 | 0.56 | 1.51 | 3.54 |
| $C_3N_3$-134 | 0.53 | 1.27 | 3.50 |
| $C_3N_3$-135 | 0.52 | 1.42 | 3.48 |
| $C_3N_3$-136 | 0.59 | 1.46 | 3.56 |
| $C_3N_3$-156 | 0.46 | 1.32 | 3.44 |
| $C_3N_3$-235 | 0.63 | 1.57 | 3.52 |
| $C_3N_3$-236 | 0.73 | 1.76 | 3.63 |
| $C_4N_2$-12 | 0.59 | 1.59 | 3.50 |
| $C_4N_2$-13 | 0.51 | 1.41 | 3.36 |
| $C_4N_2$-14 | 0.59 | 1.60 | 3.47 |
| $C_4N_2$-15 | 0.58 | 1.48 | 3.48 |
| $C_4N_2$-16 | 0.46 | 1.39 | 3.37 |
| $C_4N_2$-23 | 0.46 | 1.34 | 3.45 |
| $C_4N_2$-25 | 0.54 | 1.39 | 3.50 |
| $C_4N_2$-26 | 0.47 | 1.37 | 3.45 |
| $C_4N_2$-36 | 0.34 | 1.22 | 3.32 |
| $C_5N$-1 | 0.21 | 1.31 | 3.34 |
| $C_5N$-2 | 0.39 | 1.37 | 3.47 |
| $C_5N$-3 | 0.11 | 1.16 | 3.19 |
| $C_6$ | 0.10 | 1.06 | 3.41 |

**Table S7.** Calculated adsorption free energies of *OH, *O, and *OOH on CoNi-$C_{6-x}N_x$ DACs.

| | $\Delta G_{*OH}$ (eV) | $\Delta G_{*O}$ (eV) | $\Delta G_{*OOH}$ (eV) |
|---|---|---|---|
| $N_6$ | 1.15 | 2.56 | 4.12 |
| $CN_5$-1 | 1.07 | 2.43 | 4.05 |
| $CN_5$-2 | 1.31 | 2.91 | 4.25 |
| $CN_5$-3 | 1.17 | 2.72 | 4.11 |
| $C_2N_4$-12 | 1.38 | 2.82 | 4.27 |
| $C_2N_4$-13 | 1.28 | 2.71 | 4.23 |
| $C_2N_4$-14 | 1.19 | 2.39 | 4.15 |
| $C_2N_4$-15 | 1.25 | 2.77 | 4.16 |
| $C_2N_4$-16 | 1.31 | 2.75 | 4.25 |
| $C_2N_4$-23 | 1.30 | 3.04 | 4.24 |
| $C_2N_4$-25 | 1.26 | 2.89 | 4.21 |
| $C_2N_4$-26 | 1.34 | 3.12 | 4.28 |
| $C_2N_4$-36 | 1.39 | 3.14 | 4.31 |
| $C_3N_3$-123 | 1.31 | 2.86 | 4.30 |
| $C_3N_3$-124 | 1.26 | 2.63 | 4.17 |
| $C_3N_3$-125 | 1.15 | 2.61 | 4.11 |
| $C_3N_3$-126 | 1.31 | 2.86 | 4.29 |
| $C_3N_3$-134 | 1.21 | 2.63 | 4.16 |
| $C_3N_3$-135 | 1.33 | 2.79 | 4.21 |
| $C_3N_3$-136 | 1.33 | 2.84 | 4.21 |
| $C_3N_3$-156 | 1.29 | 2.73 | 4.19 |
| $C_3N_3$-235 | 1.21 | 2.83 | 4.18 |
| $C_3N_3$-236 | 1.47 | 3.04 | 4.31 |
| $C_4N_2$-12 | 1.04 | 2.59 | 3.95 |
| $C_4N_2$-13 | 0.77 | 2.27 | 3.70 |
| $C_4N_2$-14 | 0.76 | 2.41 | 3.74 |
| $C_4N_2$-15 | 1.01 | 2.47 | 3.91 |
| $C_4N_2$-16 | 0.80 | 2.26 | 3.72 |
| $C_4N_2$-23 | 0.97 | 2.37 | 3.90 |
| $C_4N_2$-25 | 1.15 | 2.53 | 4.07 |
| $C_4N_2$-26 | 0.98 | 2.35 | 3.88 |
| $C_4N_2$-36 | 0.82 | 2.17 | 3.73 |
| $C_5N$-1 | 0.69 | 1.88 | 3.67 |
| $C_5N$-2 | 0.83 | 2.12 | 3.74 |
| $C_5N$-3 | 0.66 | 1.77 | 3.68 |
| $C_6$ | 0.62 | 1.66 | 3.58 |

**Table S8.** Calculated adsorption free energies of *OH, *O, and *OOH on CoCu-$C_{6-x}N_x$ DACs.

| | $\Delta G_{*OH}$ (eV) | $\Delta G_{*O}$ (eV) | $\Delta G_{*OOH}$ (eV) |
|---|---|---|---|
| $N_6$ | 1.20 | 2.35 | 4.06 |
| $CN_5$-1 | 1.16 | 2.43 | 4.04 |
| $CN_5$-2 | 1.13 | 2.57 | 4.12 |
| $CN_5$-3 | 1.19 | 2.46 | 3.93 |
| $C_2N_4$-12 | 1.13 | 2.48 | 4.22 |
| $C_2N_4$-13 | 1.22 | 2.28 | 4.16 |
| $C_2N_4$-14 | 0.92 | 1.93 | 3.89 |
| $C_2N_4$-15 | 1.14 | 2.46 | 4.08 |
| $C_2N_4$-16 | 1.18 | 2.22 | 4.12 |
| $C_2N_4$-23 | 1.09 | 2.54 | 4.09 |
| $C_2N_4$-25 | 1.25 | 2.99 | 4.20 |
| $C_2N_4$-26 | 1.27 | 2.84 | 4.24 |
| $C_2N_4$-36 | 1.21 | 2.65 | 4.17 |
| $C_3N_3$-123 | 1.06 | 2.43 | 4.04 |
| $C_3N_3$-124 | 1.16 | 2.77 | 4.07 |
| $C_3N_3$-125 | 1.29 | 2.76 | 4.20 |
| $C_3N_3$-126 | 1.27 | 2.84 | 4.22 |
| $C_3N_3$-134 | 0.98 | 2.03 | 3.96 |
| $C_3N_3$-135 | 1.21 | 2.71 | 4.14 |
| $C_3N_3$-136 | 1.10 | 2.42 | 4.05 |
| $C_3N_3$-156 | 1.07 | 2.41 | 4.04 |
| $C_3N_3$-235 | 1.24 | 2.95 | 4.20 |
| $C_3N_3$-236 | 1.27 | 2.97 | 4.22 |
| $C_4N_2$-12 | 1.27 | 2.80 | 4.22 |
| $C_4N_2$-13 | 1.20 | 2.71 | 4.15 |
| $C_4N_2$-14 | 1.26 | 2.92 | 4.23 |
| $C_4N_2$-15 | 1.21 | 2.73 | 4.13 |
| $C_4N_2$-16 | 1.21 | 2.66 | 4.13 |
| $C_4N_2$-23 | 1.11 | 2.50 | 4.14 |
| $C_4N_2$-25 | 1.08 | 2.41 | 4.11 |
| $C_4N_2$-26 | 1.20 | 2.57 | 4.16 |
| $C_4N_2$-36 | 1.16 | 2.48 | 4.09 |
| $C_5N$-1 | 0.89 | 2.37 | 3.80 |
| $C_5N$-2 | 1.08 | 2.40 | 4.02 |
| $C_5N$-3 | 0.84 | 2.21 | 3.77 |
| $C_6$ | 0.82 | 2.07 | 3.77 |

**Table S9.** Calculated reaction free energies ($\Delta G$, eV) of OER elementary steps, OER overpotentials ($\eta_{OER}$, V), and potential-determining steps (PDSs) for $Fe_2$-$C_{6-x}N_x$ DACs.

| | $\Delta G_a$ | $\Delta G_b$ | $\Delta G_c$ | $\Delta G_d$ | $\eta_{OER}$ | PDS |
|---|---|---|---|---|---|---|
| $N_6$ | 1.06 | 1.08 | 1.75 | 1.03 | 0.52 | *O → *OOH |
| $CN_5$-1 | 0.87 | 0.87 | 2.08 | 1.10 | 0.85 | *O → *OOH |
| $CN_5$-2 | 0.79 | 0.96 | 2.03 | 1.14 | 0.80 | *O → *OOH |
| $C_2N_4$-12 | 0.64 | 0.82 | 2.17 | 1.29 | 0.94 | *O → *OOH |
| $C_2N_4$-13 | 0.79 | 0.75 | 2.10 | 1.28 | 0.87 | *O → *OOH |
| $C_2N_4$-14 | 0.68 | 0.82 | 2.02 | 1.40 | 0.79 | *O → *OOH |
| $C_2N_4$-15 | 0.57 | 0.84 | 2.11 | 1.40 | 0.88 | *O → *OOH |
| $C_2N_4$-16 | 0.68 | 0.84 | 2.11 | 1.28 | 0.88 | *O → *OOH |
| $C_2N_4$-25 | 0.50 | 0.86 | 2.09 | 1.46 | 0.86 | *O → *OOH |
| $C_3N_3$-123 | 0.62 | 0.63 | 2.23 | 1.45 | 1.00 | *O → *OOH |
| $C_3N_3$-124 | 0.54 | 0.72 | 2.09 | 1.58 | 0.86 | *O → *OOH |
| $C_3N_3$-125 | 0.57 | 0.60 | 1.98 | 1.77 | 0.75 | *O → *OOH |
| $C_3N_3$-126 | 0.69 | 0.58 | 2.21 | 1.44 | 0.98 | *O → *OOH |
| $C_3N_3$-134 | 0.58 | 0.80 | 2.18 | 1.36 | 0.95 | *O → *OOH |
| $C_3N_3$-135 | 0.58 | 0.61 | 2.17 | 1.56 | 0.94 | *O → *OOH |
| $C_4N_2$-12 | 0.54 | 0.70 | 2.13 | 1.55 | 0.90 | *O → *OOH |
| $C_4N_2$-13 | 0.11 | 0.76 | 2.29 | 1.77 | 1.06 | *O → *OOH |
| $C_4N_2$-14 | 0.09 | 0.87 | 2.14 | 1.82 | 0.91 | *O → *OOH |
| $C_4N_2$-15 | 0.41 | 0.80 | 2.14 | 1.58 | 0.91 | *O → *OOH |
| $C_4N_2$-16 | 0.09 | 0.76 | 2.29 | 1.78 | 1.06 | *O → *OOH |
| $C_4N_2$-25 | 0.53 | 0.80 | 2.13 | 1.46 | 0.90 | *O → *OOH |
| $C_5N$-1 | -0.09 | 0.73 | 2.48 | 1.80 | 1.25 | *O → *OOH |
| $C_5N$-2 | 0.17 | 0.69 | 2.32 | 1.75 | 1.09 | *O → *OOH |
| $C_6$ | -0.21 | 0.39 | 2.63 | 2.10 | 1.40 | *O → *OOH |

**Table S10.** Calculated reaction free energies ($\Delta G$, eV) of OER elementary steps, OER overpotentials ($\eta_{OER}$, V), and potential-determining steps (PDSs) for $Co_2$-$C_{6-x}N_x$ DACs.

| | $\Delta G_a$ | $\Delta G_b$ | $\Delta G_c$ | $\Delta G_d$ | $\eta_{OER}$ | PDS |
|---|---|---|---|---|---|---|
| $N_6$ | 1.36 | 1.67 | 1.24 | 0.65 | 0.44 | *OH → *O |
| $CN_5$-1 | 1.46 | 1.50 | 1.36 | 0.60 | 0.27 | *OH → *O |
| $CN_5$-2 | 1.41 | 1.71 | 1.21 | 0.59 | 0.48 | *OH → *O |
| $C_2N_4$-12 | 1.30 | 1.59 | 1.38 | 0.65 | 0.36 | *OH → *O |
| $C_2N_4$-13 | 1.42 | 1.56 | 1.31 | 0.64 | 0.33 | *OH → *O |
| $C_2N_4$-14 | 1.27 | 1.54 | 1.40 | 0.72 | 0.31 | *OH → *O |
| $C_2N_4$-15 | 1.32 | 1.48 | 1.39 | 0.73 | 0.25 | *OH → *O |
| $C_2N_4$-16 | 1.39 | 1.56 | 1.31 | 0.66 | 0.33 | *OH → *O |
| $C_2N_4$-25 | 1.10 | 1.63 | 1.32 | 0.87 | 0.40 | *OH → *O |
| $C_3N_3$-123 | 1.00 | 1.61 | 1.36 | 0.95 | 0.38 | *OH → *O |
| $C_3N_3$-124 | 0.95 | 1.42 | 1.48 | 1.07 | 0.25 | *O → *OOH |
| $C_3N_3$-125 | 0.83 | 1.33 | 1.52 | 1.24 | 0.29 | *O → *OOH |
| $C_3N_3$-126 | 1.02 | 1.55 | 1.43 | 0.93 | 0.32 | *OH → *O |
| $C_3N_3$-134 | 1.08 | 1.44 | 1.48 | 0.92 | 0.25 | *O → *OOH |
| $C_3N_3$-135 | 1.07 | 1.45 | 1.40 | 1.00 | 0.22 | *OH → *O |
| $C_4N_2$-12 | 0.95 | 1.35 | 1.59 | 1.03 | 0.36 | *O → *OOH |
| $C_4N_2$-13 | 0.51 | 1.17 | 1.81 | 1.43 | 0.58 | *O → *OOH |
| $C_4N_2$-14 | 0.69 | 1.36 | 1.61 | 1.26 | 0.38 | *O → *OOH |
| $C_4N_2$-15 | 0.92 | 1.28 | 1.63 | 1.09 | 0.40 | *O → *OOH |
| $C_4N_2$-16 | 0.56 | 1.17 | 1.80 | 1.39 | 0.57 | *O → *OOH |
| $C_4N_2$-25 | 1.00 | 1.34 | 1.56 | 1.03 | 0.33 | *O → *OOH |
| $C_5N$-1 | 0.60 | 1.02 | 1.85 | 1.45 | 0.62 | *O → *OOH |
| $C_5N$-2 | 0.64 | 1.10 | 1.86 | 1.33 | 0.63 | *O → *OOH |
| $C_6$ | 0.47 | 1.00 | 1.92 | 1.53 | 0.69 | *O → *OOH |

**Table S11.** Calculated reaction free energies (Δ$G$, eV) of OER elementary steps, OER overpotentials ($\eta_{OER}$, V), and potential-determining steps (PDSs) for FeCo-$C_{6-x}N_x$ DACs.

| | $\Delta G_a$ | $\Delta G_b$ | $\Delta G_c$ | $\Delta G_d$ | $\eta_{OER}$ | PDS |
|---|---|---|---|---|---|---|
| $N_6$ | 0.97 | 1.03 | 1.82 | 1.10 | 0.59 | *O → *OOH |
| $CN_5$-1 | 0.98 | 0.77 | 1.97 | 1.21 | 0.74 | *O → *OOH |
| $CN_5$-2 | 0.82 | 0.97 | 1.90 | 1.23 | 0.67 | *O → *OOH |
| $CN_5$-3 | 0.84 | 1.06 | 1.80 | 1.22 | 0.57 | *O → *OOH |
| $C_2N_4$-12 | 0.68 | 0.93 | 2.05 | 1.26 | 0.82 | *O → *OOH |
| $C_2N_4$-13 | 0.98 | 0.76 | 1.96 | 1.22 | 0.73 | *O → *OOH |
| $C_2N_4$-14 | 0.77 | 0.82 | 2.00 | 1.33 | 0.77 | *O → *OOH |
| $C_2N_4$-15 | 0.67 | 0.88 | 2.05 | 1.32 | 0.82 | *O → *OOH |
| $C_2N_4$-16 | 0.98 | 0.73 | 1.99 | 1.23 | 0.76 | *O → *OOH |
| $C_2N_4$-23 | 0.84 | 0.91 | 1.95 | 1.21 | 0.72 | *O → *OOH |
| $C_2N_4$-25 | 0.60 | 0.90 | 2.11 | 1.31 | 0.88 | *O → *OOH |
| $C_2N_4$-26 | 0.72 | 1.05 | 1.94 | 1.22 | 0.71 | *O → *OOH |
| $C_2N_4$-36 | 0.80 | 1.10 | 1.83 | 1.19 | 0.60 | *O → *OOH |
| $C_3N_3$-123 | 0.82 | 0.55 | 2.23 | 1.32 | 1.00 | *O → *OOH |
| $C_3N_3$-124 | 0.47 | 0.82 | 2.18 | 1.46 | 0.95 | *O → *OOH |
| $C_3N_3$-125 | 0.37 | 0.74 | 2.19 | 1.62 | 0.96 | *O → *OOH |
| $C_3N_3$-126 | 0.62 | 0.78 | 2.20 | 1.32 | 0.97 | *O → *OOH |
| $C_3N_3$-134 | 0.62 | 0.89 | 2.12 | 1.28 | 0.89 | *O → *OOH |
| $C_3N_3$-135 | 0.54 | 0.81 | 2.18 | 1.39 | 0.95 | *O → *OOH |
| $C_3N_3$-136 | 0.76 | 0.85 | 2.07 | 1.23 | 0.84 | *O → *OOH |
| $C_3N_3$-156 | 0.69 | 0.65 | 2.17 | 1.41 | 0.94 | *O → *OOH |
| $C_3N_3$-235 | 0.42 | 0.86 | 2.16 | 1.48 | 0.93 | *O → *OOH |
| $C_3N_3$-236 | 0.54 | 0.87 | 2.13 | 1.37 | 0.90 | *O → *OOH |
| $C_4N_2$-12 | 0.48 | 0.85 | 2.13 | 1.46 | 0.90 | *O → *OOH |
| $C_4N_2$-13 | 0.13 | 0.77 | 2.29 | 1.74 | 1.06 | *O → *OOH |
| $C_4N_2$-14 | 0.36 | 0.76 | 2.15 | 1.65 | 0.92 | *O → *OOH |
| $C_4N_2$-15 | 0.48 | 0.77 | 2.20 | 1.48 | 0.97 | *O → *OOH |
| $C_4N_2$-16 | 0.14 | 0.77 | 2.31 | 1.71 | 1.08 | *O → *OOH |
| $C_4N_2$-23 | 0.39 | 0.77 | 2.27 | 1.48 | 1.04 | *O → *OOH |
| $C_4N_2$-25 | 0.58 | 0.91 | 2.14 | 1.30 | 0.91 | *O → *OOH |
| $C_4N_2$-26 | 0.38 | 0.78 | 2.28 | 1.47 | 1.05 | *O → *OOH |
| $C_4N_2$-36 | 0.00 | 0.66 | 2.50 | 1.75 | 1.27 | *O → *OOH |
| $C_5N$-1 | -0.03 | 0.76 | 2.45 | 1.74 | 1.22 | *O → *OOH |
| $C_5N$-2 | 0.32 | 0.75 | 2.29 | 1.56 | 1.06 | *O → *OOH |
| $C_5N$-3 | -0.08 | 0.57 | 2.60 | 1.82 | 1.37 | *O → *OOH |
| $C_6$ | -0.09 | 0.48 | 2.68 | 1.85 | 1.45 | *O → *OOH |

**Table S12.** Calculated reaction free energies (Δ$G$, eV) of OER elementary steps, OER overpotentials ($\eta_{OER}$, V), and potential-determining steps (PDSs) for FeNi-$C_{6-x}N_x$ DACs.

| | $\Delta G_a$ | $\Delta G_b$ | $\Delta G_c$ | $\Delta G_d$ | $\eta_{OER}$ | PDS |
|---|---|---|---|---|---|---|
| $N_6$ | 0.70 | 0.93 | 1.96 | 1.32 | 0.73 | *O → *OOH |
| $CN_5$-1 | 0.65 | 0.78 | 2.15 | 1.34 | 0.92 | *O → *OOH |
| $CN_5$-2 | 0.73 | 1.00 | 1.90 | 1.29 | 0.67 | *O → *OOH |
| $CN_5$-3 | 0.82 | 0.93 | 1.96 | 1.21 | 0.73 | *O → *OOH |
| $C_2N_4$-12 | 0.64 | 0.97 | 2.01 | 1.31 | 0.78 | *O → *OOH |
| $C_2N_4$-13 | 0.68 | 0.97 | 1.99 | 1.28 | 0.76 | *O → *OOH |
| $C_2N_4$-14 | 0.68 | 0.75 | 2.12 | 1.36 | 0.89 | *O → *OOH |
| $C_2N_4$-15 | 0.59 | 0.86 | 2.08 | 1.38 | 0.85 | *O → *OOH |
| $C_2N_4$-16 | 0.88 | 0.76 | 1.99 | 1.29 | 0.76 | *O → *OOH |
| $C_2N_4$-23 | 0.73 | 0.98 | 1.91 | 1.31 | 0.68 | *O → *OOH |
| $C_2N_4$-25 | 0.64 | 0.94 | 1.92 | 1.41 | 0.69 | *O → *OOH |
| $C_2N_4$-26 | 0.73 | 0.99 | 1.90 | 1.31 | 0.67 | *O → *OOH |
| $C_2N_4$-36 | 0.84 | 1.05 | 1.82 | 1.21 | 0.59 | *O → *OOH |
| $C_3N_3$-123 | 0.77 | 0.89 | 1.96 | 1.31 | 0.73 | *O → *OOH |
| $C_3N_3$-124 | 0.56 | 0.83 | 2.04 | 1.49 | 0.81 | *O → *OOH |
| $C_3N_3$-125 | 0.53 | 0.80 | 2.10 | 1.48 | 0.87 | *O → *OOH |
| $C_3N_3$-126 | 0.77 | 0.85 | 1.98 | 1.33 | 0.75 | *O → *OOH |
| $C_3N_3$-134 | 0.66 | 0.84 | 2.05 | 1.37 | 0.82 | *O → *OOH |
| $C_3N_3$-135 | 0.72 | 0.82 | 2.02 | 1.37 | 0.79 | *O → *OOH |
| $C_3N_3$-136 | 0.84 | 0.82 | 2.01 | 1.26 | 0.78 | *O → *OOH |
| $C_3N_3$-156 | 0.67 | 0.85 | 2.05 | 1.35 | 0.82 | *O → *OOH |
| $C_3N_3$-235 | 0.55 | 1.03 | 1.97 | 1.38 | 0.74 | *O → *OOH |
| $C_3N_3$-236 | 0.87 | 0.88 | 1.90 | 1.28 | 0.67 | *O → *OOH |
| $C_4N_2$-12 | 0.47 | 1.00 | 2.09 | 1.37 | 0.86 | *O → *OOH |
| $C_4N_2$-13 | 0.18 | 1.00 | 2.11 | 1.63 | 0.88 | *O → *OOH |
| $C_4N_2$-14 | 0.32 | 0.92 | 2.19 | 1.49 | 0.96 | *O → *OOH |
| $C_4N_2$-15 | 0.48 | 0.92 | 2.12 | 1.39 | 0.89 | *O → *OOH |
| $C_4N_2$-16 | 0.19 | 1.00 | 2.11 | 1.61 | 0.88 | *O → *OOH |
| $C_4N_2$-23 | 0.37 | 0.96 | 2.21 | 1.38 | 0.98 | *O → *OOH |
| $C_4N_2$-25 | 0.61 | 0.90 | 2.07 | 1.34 | 0.84 | *O → *OOH |
| $C_4N_2$-26 | 0.36 | 0.99 | 2.16 | 1.41 | 0.93 | *O → *OOH |
| $C_4N_2$-36 | 0.15 | 0.93 | 2.08 | 1.75 | 0.85 | *O → *OOH |
| $C_5N$-1 | 0.13 | 0.83 | 2.39 | 1.57 | 1.16 | *O → *OOH |
| $C_5N$-2 | 0.33 | 0.93 | 2.20 | 1.46 | 0.97 | *O → *OOH |
| $C_5N$-3 | 0.08 | 0.75 | 2.46 | 1.64 | 1.23 | *O → *OOH |
| $C_6$ | 0.06 | 0.57 | 2.59 | 1.70 | 1.36 | *O → *OOH |

**Table S13.** Calculated reaction free energies (Δ$G$, eV) of OER elementary steps, OER overpotentials ($\eta_{OER}$, V), and potential-determining steps (PDSs) for FeCu-$C_{6-x}N_x$ DACs.

| | $\Delta G_a$ | $\Delta G_b$ | $\Delta G_c$ | $\Delta G_d$ | $\eta_{OER}$ | PDS |
|---|---|---|---|---|---|---|
| $N_6$ | 0.50 | 0.92 | 2.13 | 1.37 | 0.90 | *O → *OOH |
| $CN_5$-1 | 0.42 | 1.04 | 1.93 | 1.53 | 0.70 | *O → *OOH |
| $CN_5$-2 | 0.51 | 1.00 | 1.91 | 1.50 | 0.68 | *O → *OOH |
| $CN_5$-3 | 1.03 | 0.66 | 1.99 | 1.24 | 0.76 | *O → *OOH |
| $C_2N_4$-12 | 0.54 | 0.94 | 2.10 | 1.34 | 0.87 | *O → *OOH |
| $C_2N_4$-13 | 0.73 | 0.74 | 2.11 | 1.34 | 0.88 | *O → *OOH |
| $C_2N_4$-14 | 0.58 | 0.83 | 2.11 | 1.40 | 0.88 | *O → *OOH |
| $C_2N_4$-15 | 0.43 | 0.86 | 2.15 | 1.48 | 0.92 | *O → *OOH |
| $C_2N_4$-16 | 0.67 | 0.95 | 1.94 | 1.35 | 0.71 | *O → *OOH |
| $C_2N_4$-23 | 0.60 | 0.89 | 2.01 | 1.42 | 0.78 | *O → *OOH |
| $C_2N_4$-25 | 0.65 | 0.96 | 1.94 | 1.38 | 0.71 | *O → *OOH |
| $C_2N_4$-26 | 0.80 | 1.01 | 1.89 | 1.22 | 0.66 | *O → *OOH |
| $C_2N_4$-36 | 0.69 | 0.96 | 1.94 | 1.33 | 0.71 | *O → *OOH |
| $C_3N_3$-123 | 0.45 | 0.99 | 2.04 | 1.44 | 0.81 | *O → *OOH |
| $C_3N_3$-124 | 0.44 | 0.79 | 2.16 | 1.53 | 0.93 | *O → *OOH |
| $C_3N_3$-125 | 0.52 | 0.88 | 2.07 | 1.45 | 0.84 | *O → *OOH |
| $C_3N_3$-126 | 0.56 | 0.96 | 2.02 | 1.38 | 0.79 | *O → *OOH |
| $C_3N_3$-134 | 0.53 | 0.74 | 2.23 | 1.42 | 1.00 | *O → *OOH |
| $C_3N_3$-135 | 0.52 | 0.90 | 2.07 | 1.44 | 0.84 | *O → *OOH |
| $C_3N_3$-136 | 0.59 | 0.87 | 2.09 | 1.36 | 0.86 | *O → *OOH |
| $C_3N_3$-156 | 0.46 | 0.86 | 2.12 | 1.48 | 0.89 | *O → *OOH |
| $C_3N_3$-235 | 0.63 | 0.95 | 1.94 | 1.40 | 0.71 | *O → *OOH |
| $C_3N_3$-236 | 0.73 | 1.03 | 1.87 | 1.29 | 0.64 | *O → *OOH |
| $C_4N_2$-12 | 0.59 | 1.01 | 1.91 | 1.42 | 0.68 | *O → *OOH |
| $C_4N_2$-13 | 0.51 | 0.90 | 1.96 | 1.56 | 0.73 | *O → *OOH |
| $C_4N_2$-14 | 0.59 | 1.01 | 1.87 | 1.45 | 0.64 | *O → *OOH |
| $C_4N_2$-15 | 0.58 | 0.90 | 2.00 | 1.44 | 0.77 | *O → *OOH |
| $C_4N_2$-16 | 0.46 | 0.93 | 1.98 | 1.55 | 0.75 | *O → *OOH |
| $C_4N_2$-23 | 0.46 | 0.88 | 2.11 | 1.47 | 0.88 | *O → *OOH |
| $C_4N_2$-25 | 0.54 | 0.84 | 2.12 | 1.42 | 0.89 | *O → *OOH |
| $C_4N_2$-26 | 0.47 | 0.90 | 2.07 | 1.47 | 0.84 | *O → *OOH |
| $C_4N_2$-36 | 0.34 | 0.88 | 2.10 | 1.60 | 0.87 | *O → *OOH |
| $C_5N$-1 | 0.21 | 1.10 | 2.03 | 1.58 | 0.80 | *O → *OOH |
| $C_5N$-2 | 0.39 | 0.98 | 2.09 | 1.45 | 0.86 | *O → *OOH |
| $C_5N$-3 | 0.11 | 1.05 | 2.03 | 1.73 | 0.80 | *O → *OOH |
| $C_6$ | 0.10 | 0.95 | 2.35 | 1.51 | 1.12 | *O → *OOH |

**Table S14.** Calculated reaction free energies (Δ*G*, eV) of OER elementary steps, OER overpotentials ($\eta_{OER}$, V), and potential-determining steps (PDSs) for CoNi-$C_{6-x}N_x$ DACs.

| | $\Delta G_a$ | $\Delta G_b$ | $\Delta G_c$ | $\Delta G_d$ | $\eta_{OER}$ | PDS |
|---|---|---|---|---|---|---|
| $N_6$ | 1.15 | 1.41 | 1.56 | 0.80 | 0.33 | *O → *OOH |
| $CN_5$-1 | 1.07 | 1.36 | 1.62 | 0.87 | 0.39 | *O → *OOH |
| $CN_5$-2 | 1.31 | 1.60 | 1.33 | 0.67 | 0.37 | *OH → *O |
| $CN_5$-3 | 1.17 | 1.55 | 1.38 | 0.81 | 0.32 | *OH → *O |
| $C_2N_4$-12 | 1.38 | 1.44 | 1.45 | 0.65 | 0.22 | *O → *OOH |
| $C_2N_4$-13 | 1.28 | 1.43 | 1.51 | 0.69 | 0.28 | *O → *OOH |
| $C_2N_4$-14 | 1.19 | 1.20 | 1.77 | 0.77 | 0.54 | *O → *OOH |
| $C_2N_4$-15 | 1.25 | 1.52 | 1.39 | 0.76 | 0.29 | *OH → *O |
| $C_2N_4$-16 | 1.31 | 1.43 | 1.51 | 0.67 | 0.28 | *O → *OOH |
| $C_2N_4$-23 | 1.30 | 1.74 | 1.20 | 0.68 | 0.51 | *OH → *O |
| $C_2N_4$-25 | 1.26 | 1.63 | 1.32 | 0.71 | 0.40 | *OH → *O |
| $C_2N_4$-26 | 1.34 | 1.78 | 1.15 | 0.64 | 0.55 | *OH → *O |
| $C_2N_4$-36 | 1.39 | 1.75 | 1.17 | 0.61 | 0.52 | *OH → *O |
| $C_3N_3$-123 | 1.31 | 1.55 | 1.44 | 0.62 | 0.32 | *OH → *O |
| $C_3N_3$-124 | 1.26 | 1.37 | 1.54 | 0.75 | 0.31 | *O → *OOH |
| $C_3N_3$-125 | 1.15 | 1.46 | 1.50 | 0.81 | 0.27 | *O → *OOH |
| $C_3N_3$-126 | 1.31 | 1.54 | 1.43 | 0.63 | 0.31 | *OH → *O |
| $C_3N_3$-134 | 1.21 | 1.42 | 1.53 | 0.76 | 0.30 | *O → *OOH |
| $C_3N_3$-135 | 1.33 | 1.46 | 1.42 | 0.71 | 0.23 | *OH → *O |
| $C_3N_3$-136 | 1.33 | 1.50 | 1.38 | 0.71 | 0.27 | *OH → *O |
| $C_3N_3$-156 | 1.29 | 1.44 | 1.46 | 0.73 | 0.23 | *O → *OOH |
| $C_3N_3$-235 | 1.21 | 1.61 | 1.35 | 0.74 | 0.38 | *OH → *O |
| $C_3N_3$-236 | 1.47 | 1.57 | 1.27 | 0.61 | 0.34 | *OH → *O |
| $C_4N_2$-12 | 1.04 | 1.54 | 1.37 | 0.97 | 0.31 | *OH → *O |
| $C_4N_2$-13 | 0.77 | 1.50 | 1.42 | 1.22 | 0.27 | *OH → *O |
| $C_4N_2$-14 | 0.76 | 1.65 | 1.33 | 1.18 | 0.42 | *OH → *O |
| $C_4N_2$-15 | 1.01 | 1.47 | 1.44 | 1.01 | 0.24 | *OH → *O |
| $C_4N_2$-16 | 0.80 | 1.46 | 1.46 | 1.20 | 0.23 | *O → *OOH |
| $C_4N_2$-23 | 0.97 | 1.40 | 1.53 | 1.02 | 0.30 | *O → *OOH |
| $C_4N_2$-25 | 1.15 | 1.38 | 1.55 | 0.85 | 0.32 | *O → *OOH |
| $C_4N_2$-26 | 0.98 | 1.38 | 1.53 | 1.04 | 0.30 | *O → *OOH |
| $C_4N_2$-36 | 0.82 | 1.35 | 1.56 | 1.19 | 0.33 | *O → *OOH |
| $C_5N$-1 | 0.69 | 1.18 | 1.79 | 1.25 | 0.56 | *O → *OOH |
| $C_5N$-2 | 0.83 | 1.29 | 1.62 | 1.18 | 0.39 | *O → *OOH |
| $C_5N$-3 | 0.66 | 1.10 | 1.91 | 1.24 | 0.68 | *O → *OOH |
| $C_6$ | 0.62 | 1.04 | 1.92 | 1.34 | 0.69 | *O → *OOH |

**Table S15.** Calculated reaction free energies ($\Delta G$, eV) of OER elementary steps, OER overpotentials ($\eta_{OER}$, V), and potential-determining steps (PDSs) for CoCu-$C_{6-x}N_x$ DACs.

| | $\Delta G_a$ | $\Delta G_b$ | $\Delta G_c$ | $\Delta G_d$ | $\eta_{OER}$ | PDS |
|---|---|---|---|---|---|---|
| $N_6$ | 1.20 | 1.15 | 1.71 | 0.86 | 0.48 | *O → *OOH |
| $CN_5$-1 | 1.16 | 1.27 | 1.61 | 0.88 | 0.38 | *O → *OOH |
| $CN_5$-2 | 1.13 | 1.44 | 1.56 | 0.80 | 0.33 | *O → *OOH |
| $CN_5$-3 | 1.19 | 1.27 | 1.47 | 0.99 | 0.24 | *O → *OOH |
| $C_2N_4$-12 | 1.13 | 1.35 | 1.74 | 0.70 | 0.51 | *O → *OOH |
| $C_2N_4$-13 | 1.22 | 1.06 | 1.88 | 0.76 | 0.65 | *O → *OOH |
| $C_2N_4$-14 | 0.92 | 1.01 | 1.96 | 1.03 | 0.73 | *O → *OOH |
| $C_2N_4$-15 | 1.14 | 1.32 | 1.61 | 0.84 | 0.38 | *O → *OOH |
| $C_2N_4$-16 | 1.18 | 1.04 | 1.90 | 0.80 | 0.67 | *O → *OOH |
| $C_2N_4$-23 | 1.09 | 1.45 | 1.54 | 0.83 | 0.31 | *O → *OOH |
| $C_2N_4$-25 | 1.25 | 1.73 | 1.22 | 0.72 | 0.50 | *OH → *O |
| $C_2N_4$-26 | 1.27 | 1.57 | 1.39 | 0.68 | 0.34 | *OH → *O |
| $C_2N_4$-36 | 1.21 | 1.44 | 1.52 | 0.75 | 0.29 | *O → *OOH |
| $C_3N_3$-123 | 1.06 | 1.37 | 1.61 | 0.88 | 0.38 | *O → *OOH |
| $C_3N_3$-124 | 1.16 | 1.61 | 1.30 | 0.85 | 0.38 | *OH → *O |
| $C_3N_3$-125 | 1.29 | 1.47 | 1.44 | 0.72 | 0.24 | *OH → *O |
| $C_3N_3$-126 | 1.27 | 1.57 | 1.38 | 0.70 | 0.34 | *OH → *O |
| $C_3N_3$-134 | 0.98 | 1.05 | 1.93 | 0.96 | 0.70 | *O → *OOH |
| $C_3N_3$-135 | 1.21 | 1.50 | 1.43 | 0.78 | 0.27 | *OH → *O |
| $C_3N_3$-136 | 1.10 | 1.33 | 1.62 | 0.87 | 0.39 | *O → *OOH |
| $C_3N_3$-156 | 1.07 | 1.34 | 1.63 | 0.88 | 0.40 | *O → *OOH |
| $C_3N_3$-235 | 1.24 | 1.71 | 1.25 | 0.72 | 0.48 | *OH → *O |
| $C_3N_3$-236 | 1.27 | 1.71 | 1.25 | 0.70 | 0.48 | *OH → *O |
| $C_4N_2$-12 | 1.27 | 1.53 | 1.42 | 0.70 | 0.30 | *OH → *O |
| $C_4N_2$-13 | 1.20 | 1.51 | 1.44 | 0.77 | 0.28 | *OH → *O |
| $C_4N_2$-14 | 1.26 | 1.66 | 1.31 | 0.69 | 0.43 | *OH → *O |
| $C_4N_2$-15 | 1.21 | 1.52 | 1.40 | 0.79 | 0.29 | *OH → *O |
| $C_4N_2$-16 | 1.21 | 1.45 | 1.47 | 0.79 | 0.24 | *O → *OOH |
| $C_4N_2$-23 | 1.11 | 1.40 | 1.64 | 0.78 | 0.41 | *O → *OOH |
| $C_4N_2$-25 | 1.08 | 1.32 | 1.71 | 0.81 | 0.48 | *O → *OOH |
| $C_4N_2$-26 | 1.20 | 1.37 | 1.59 | 0.76 | 0.36 | *O → *OOH |
| $C_4N_2$-36 | 1.16 | 1.32 | 1.61 | 0.83 | 0.38 | *O → *OOH |
| $C_5N$-1 | 0.89 | 1.49 | 1.43 | 1.12 | 0.26 | *OH → *O |
| $C_5N$-2 | 1.08 | 1.33 | 1.62 | 0.90 | 0.39 | *O → *OOH |
| $C_5N$-3 | 0.84 | 1.37 | 1.56 | 1.15 | 0.33 | *O → *OOH |
| $C_6$ | 0.82 | 1.26 | 1.69 | 1.15 | 0.46 | *O → *OOH |

**Table S16.** Summary of features employed in the construction of machine learning models.

| Features | Description |
|---|---|
| $M_1$ | Spin magnetic moment of the $TM_1$ atom |
| $Q_1$, $Q_2$ | Net charges of the $TM_1$ and $TM_2$ atoms |
| $EN_1$, $EN_2$ | Pauling electronegativity of the $TM_1$ and $TM_2$ atom |
| $corrd_1$, $corrd_2$, $corrd_3$, $corrd_4$, $corrd_5$, $corrd_6$, | Binary encoding of nonmetal atom sites bonded to the TM center (C = 1, N = 0). |
| $N_{d1}$, $N_{d2}$ | Valence *d*-electron counts of $TM_1$ and $TM_2$ atoms |
| $\Delta EN_1$, $\Delta EN_2$ | The difference between the average electronegativity of the nonmetal atoms bonded to the $TM_1/TM_2$ atoms and that of the $TM_1/TM_2$ atoms. |
| $N_{C1}$, $N_{C2}$ | Coordination numbers of C atoms around $TM_1$ and $TM_2$ |
| $N_{N1}$, $N_{N2}$ | Coordination numbers of N atoms around $TM_1$ and $TM_2$ |

**Table S17.** Performance indicators of the ML models, including the coefficient of determination ($R^2$), root mean square error (RMSE), and mean absolute error (MAE) of $\eta_{OER}$

| | $R^2$ | RMSE (in V) | MAE (in V) |
|---|---|---|---|
| Train set | 0.980 | 0.040 | 0.030 |
| Test set | 0.910 | 0.079 | 0.057 |

**Table S18.** Calculated adsorption free energies of *OH, *O, and *OOH on CoNi-$C_{6-x}N_x$ DACs at 0 $V_{RHE}$ under constant-potential conditions.

| | $\Delta G_{*OH}$ (eV) | $\Delta G_{*O}$ (eV) | $\Delta G_{*OOH}$ (eV) |
|---|---|---|---|
| $N_6$ | 1.35 | 2.83 | 4.25 |
| $CN_5$-1 | 1.22 | 2.79 | 4.16 |
| $CN_5$-2 | 1.42 | 3.14 | 4.32 |
| $CN_5$-3 | 1.29 | 2.95 | 4.18 |
| $C_2N_4$-12 | 1.40 | 3.04 | 4.35 |
| $C_2N_4$-13 | 1.34 | 2.90 | 4.28 |
| $C_2N_4$-15 | 1.35 | 2.97 | 4.25 |
| $C_2N_4$-16 | 1.33 | 2.92 | 4.27 |
| $C_2N_4$-25 | 1.38 | 3.10 | 4.29 |
| $C_3N_3$-123 | 1.31 | 2.93 | 4.26 |
| $C_3N_3$-124 | 1.23 | 2.71 | 4.15 |
| $C_3N_3$-125 | 1.04 | 2.66 | 3.96 |
| $C_3N_3$-126 | 1.33 | 2.95 | 4.26 |
| $C_3N_3$-134 | 1.29 | 2.77 | 4.23 |
| $C_3N_3$-135 | 1.32 | 2.87 | 4.19 |
| $C_3N_3$-136 | 1.38 | 2.99 | 4.26 |
| $C_3N_3$-156 | 1.30 | 2.84 | 4.19 |
| $C_3N_3$-235 | 1.14 | 3.01 | 4.04 |
| $C_3N_3$-236 | 1.44 | 3.10 | 4.28 |
| $C_4N_2$-12 | 1.16 | 2.75 | 4.06 |
| $C_4N_2$-13 | 0.80 | 2.35 | 3.74 |
| $C_4N_2$-15 | 1.10 | 2.65 | 4.01 |
| $C_4N_2$-16 | 0.83 | 2.36 | 3.76 |
| $C_4N_2$-23 | 1.04 | 2.51 | 3.97 |
| $C_4N_2$-25 | 1.24 | 2.71 | 4.18 |
| $C_4N_2$-26 | 1.05 | 2.50 | 3.96 |
| $C_4N_2$-36 | 0.79 | 2.21 | 3.73 |
| $C_5N$-2 | 0.89 | 2.32 | 3.81 |

**Table S19.** Calculated adsorption free energies of *OH, *O, and *OOH on CoCu-$C_{6-x}N_x$ DACs at 0 $V_{RHE}$ under constant-potential conditions.

| | $\Delta G_{*OH}$ (eV) | $\Delta G_{*O}$ (eV) | $\Delta G_{*OOH}$ (eV) |
|---|---|---|---|
| $N_6$ | 1.26 | 2.59 | 4.15 |
| $CN_5$-1 | 1.12 | 2.30 | 4.10 |
| $CN_5$-2 | 1.24 | 2.80 | 4.20 |
| $CN_5$-3 | 0.85 | 2.34 | 3.81 |
| $C_2N_4$-15 | 1.23 | 2.67 | 4.15 |
| $C_2N_4$-23 | 1.17 | 2.72 | 4.12 |
| $C_2N_4$-26 | 1.36 | 3.02 | 4.29 |
| $C_2N_4$-36 | 1.27 | 2.98 | 4.17 |
| $C_3N_3$-123 | 1.19 | 2.67 | 4.20 |
| $C_3N_3$-124 | 1.28 | 2.53 | 4.15 |
| $C_3N_3$-125 | 1.36 | 2.98 | 4.27 |
| $C_3N_3$-126 | 1.35 | 3.02 | 4.28 |
| $C_3N_3$-135 | 1.27 | 2.87 | 4.18 |
| $C_3N_3$-136 | 1.16 | 2.57 | 4.10 |
| $C_3N_3$-156 | 1.18 | 2.62 | 4.13 |
| $C_4N_2$-12 | 1.31 | 2.91 | 4.25 |
| $C_4N_2$-13 | 1.13 | 2.74 | 4.07 |
| $C_4N_2$-15 | 1.26 | 2.84 | 4.20 |
| $C_4N_2$-16 | 1.15 | 2.72 | 4.07 |
| $C_4N_2$-23 | 1.16 | 2.60 | 4.16 |
| $C_4N_2$-26 | 1.23 | 2.68 | 4.17 |
| $C_4N_2$-36 | 1.07 | 2.54 | 4.00 |
| $C_5N$-1 | 0.93 | 2.52 | 3.87 |
| $C_5N$-2 | 1.15 | 2.55 | 4.10 |
| $C_5N$-3 | 0.85 | 2.34 | 3.81 |

**Table S20.** Calculated adsorption free energies of *OH, *O, and *OOH on $Co_2$-$C_{6-x}N_x$ DACs at at 0 $V_{RHE}$ under constant-potential conditions.

| | $\Delta G_{*OH}$ (eV) | $\Delta G_{*O}$ (eV) | $\Delta G_{*OOH}$ (eV) |
|---|---|---|---|
| $N_6$ | 1.47 | 3.27 | 4.33 |
| $CN_5$-1 | 1.50 | 3.15 | 4.35 |
| $C_2N_4$-12 | 1.28 | 3.09 | 4.20 |
| $C_2N_4$-13 | 1.44 | 3.08 | 4.29 |
| $C_2N_4$-14 | 1.32 | 2.93 | 4.24 |
| $C_2N_4$-15 | 1.29 | 2.87 | 4.14 |
| $C_2N_4$-16 | 1.44 | 3.08 | 4.30 |
| $C_2N_4$-25 | 1.06 | 2.77 | 3.95 |
| $C_3N_3$-123 | 1.04 | 2.71 | 3.98 |
| $C_3N_3$-124 | 0.95 | 2.45 | 3.87 |
| $C_3N_3$-125 | 0.82 | 2.20 | 3.70 |
| $C_3N_3$-126 | 1.04 | 2.69 | 3.98 |
| $C_3N_3$-134 | 1.17 | 2.73 | 4.08 |
| $C_3N_3$-135 | 1.07 | 2.60 | 3.93 |
| $C_4N_2$-12 | 0.97 | 2.43 | 3.91 |
| $C_4N_2$-14 | 0.82 | 2.18 | 3.74 |
| $C_4N_2$-15 | 0.95 | 2.34 | 3.87 |
| $C_4N_2$-25 | 1.02 | 2.52 | 3.94 |

**Table S21.** Calculated reaction free energies ($\Delta G$, eV) of OER elementary steps, OER overpotentials ($\eta_{OER}$, V), and potential-determining steps (PDSs) for CoNi-$C_{6-x}N_x$ DACs at 0 $V_{RHE}$ under constant-potential conditions.

| | $\Delta G_a$ | $\Delta G_b$ | $\Delta G_c$ | $\Delta G_d$ | $\eta_{OER}$ | PDS |
|---|---|---|---|---|---|---|
| $N_6$ | 1.35 | 1.48 | 1.43 | 0.67 | 0.25 | *OH → *O |
| $CN_5$-1 | 1.22 | 1.57 | 1.37 | 0.76 | 0.34 | *OH → *O |
| $CN_5$-2 | 1.42 | 1.73 | 1.17 | 0.60 | 0.50 | *OH → *O |
| $CN_5$-3 | 1.29 | 1.66 | 1.23 | 0.74 | 0.43 | *OH → *O |
| $C_2N_4$-12 | 1.40 | 1.64 | 1.31 | 0.57 | 0.41 | *OH → *O |
| $C_2N_4$-13 | 1.34 | 1.57 | 1.38 | 0.64 | 0.34 | *OH → *O |
| $C_2N_4$-15 | 1.35 | 1.61 | 1.29 | 0.67 | 0.38 | *OH → *O |
| $C_2N_4$-16 | 1.33 | 1.59 | 1.35 | 0.65 | 0.36 | *OH → *O |
| $C_2N_4$-25 | 1.38 | 1.72 | 1.19 | 0.63 | 0.49 | *OH → *O |
| $C_3N_3$-123 | 1.31 | 1.62 | 1.33 | 0.66 | 0.39 | *OH → *O |
| $C_3N_3$-124 | 1.23 | 1.48 | 1.44 | 0.77 | 0.25 | *OH → *O |
| $C_3N_3$-125 | 1.04 | 1.63 | 1.30 | 0.96 | 0.40 | *OH → *O |
| $C_3N_3$-126 | 1.33 | 1.63 | 1.31 | 0.66 | 0.40 | *OH → *O |
| $C_3N_3$-134 | 1.29 | 1.47 | 1.47 | 0.69 | 0.24 | *OH → *O |
| $C_3N_3$-135 | 1.32 | 1.56 | 1.32 | 0.73 | 0.33 | *OH → *O |
| $C_3N_3$-136 | 1.38 | 1.61 | 1.28 | 0.66 | 0.38 | *OH → *O |
| $C_3N_3$-156 | 1.30 | 1.54 | 1.35 | 0.73 | 0.31 | *OH → *O |
| $C_3N_3$-235 | 1.14 | 1.87 | 1.03 | 0.88 | 0.64 | *OH → *O |
| $C_3N_3$-236 | 1.44 | 1.66 | 1.17 | 0.64 | 0.43 | *OH → *O |
| $C_4N_2$-12 | 1.16 | 1.59 | 1.32 | 0.86 | 0.36 | *OH → *O |
| $C_4N_2$-13 | 0.80 | 1.55 | 1.39 | 1.18 | 0.32 | *OH → *O |
| $C_4N_2$-15 | 1.10 | 1.55 | 1.36 | 0.91 | 0.32 | *OH → *O |
| $C_4N_2$-16 | 0.83 | 1.53 | 1.41 | 1.16 | 0.30 | *OH → *O |
| $C_4N_2$-23 | 1.04 | 1.47 | 1.46 | 0.95 | 0.24 | *OH → *O |
| $C_4N_2$-25 | 1.24 | 1.46 | 1.47 | 0.74 | 0.24 | *O → *OOH |
| $C_4N_2$-26 | 1.05 | 1.46 | 1.46 | 0.96 | 0.23 | *O → *OOH |
| $C_4N_2$-36 | 0.79 | 1.42 | 1.51 | 1.19 | 0.28 | *O → *OOH |
| $C_5N$-2 | 0.89 | 1.44 | 1.49 | 1.11 | 0.26 | *O → *OOH |

**Table S22.** Calculated reaction free energies ($\Delta G$, eV) of OER elementary steps, OER overpotentials ($\eta_{OER}$, V), and potential-determining steps (PDSs) for CoCu-$C_{6-x}N_x$ DACs at 0 $V_{RHE}$ under constant-potential conditions.

| | $\Delta G_a$ | $\Delta G_b$ | $\Delta G_c$ | $\Delta G_d$ | $\eta_{OER}$ | PDS |
|---|---|---|---|---|---|---|
| $N_6$ | 1.26 | 1.33 | 1.56 | 0.77 | 0.33 | *O → *OOH |
| $CN_5$-1 | 1.12 | 1.17 | 1.81 | 0.82 | 0.58 | *O → *OOH |
| $CN_5$-2 | 1.24 | 1.56 | 1.40 | 0.72 | 0.33 | *OH → *O |
| $CN_5$-3 | 0.85 | 1.48 | 1.47 | 1.11 | 0.25 | *OH → *O |
| $C_2N_4$-15 | 1.23 | 1.45 | 1.48 | 0.77 | 0.25 | *O → *OOH |
| $C_2N_4$-23 | 1.17 | 1.54 | 1.40 | 0.80 | 0.31 | *OH → *O |
| $C_2N_4$-26 | 1.36 | 1.66 | 1.27 | 0.63 | 0.43 | *OH → *O |
| $C_2N_4$-36 | 1.27 | 1.71 | 1.20 | 0.75 | 0.48 | *OH → *O |
| $C_3N_3$-123 | 1.19 | 1.48 | 1.52 | 0.72 | 0.29 | *O → *OOH |
| $C_3N_3$-124 | 1.28 | 1.25 | 1.62 | 0.77 | 0.39 | *O → *OOH |
| $C_3N_3$-125 | 1.36 | 1.62 | 1.28 | 0.65 | 0.39 | *OH → *O |
| $C_3N_3$-126 | 1.35 | 1.66 | 1.27 | 0.64 | 0.43 | *OH → *O |
| $C_3N_3$-135 | 1.27 | 1.60 | 1.31 | 0.74 | 0.37 | *OH → *O |
| $C_3N_3$-136 | 1.16 | 1.41 | 1.53 | 0.82 | 0.30 | *O → *OOH |
| $C_3N_3$-156 | 1.18 | 1.44 | 1.51 | 0.79 | 0.28 | *O → *OOH |
| $C_4N_2$-12 | 1.31 | 1.60 | 1.34 | 0.67 | 0.37 | *OH → *O |
| $C_4N_2$-13 | 1.13 | 1.60 | 1.33 | 0.85 | 0.37 | *OH → *O |
| $C_4N_2$-15 | 1.26 | 1.58 | 1.36 | 0.72 | 0.35 | *OH → *O |
| $C_4N_2$-16 | 1.15 | 1.57 | 1.35 | 0.85 | 0.34 | *OH → *O |
| $C_4N_2$-23 | 1.16 | 1.45 | 1.56 | 0.76 | 0.33 | *O → *OOH |
| $C_4N_2$-26 | 1.23 | 1.45 | 1.49 | 0.75 | 0.26 | *O → *OOH |
| $C_4N_2$-36 | 1.07 | 1.47 | 1.46 | 0.92 | 0.24 | *OH → *O |
| $C_5N$-1 | 0.93 | 1.59 | 1.35 | 1.05 | 0.36 | *OH → *O |
| $C_5N$-2 | 1.15 | 1.40 | 1.55 | 0.82 | 0.32 | *O → *OOH |
| $C_5N$-3 | 0.85 | 1.48 | 1.47 | 1.11 | 0.25 | *OH → *O |

**Table S23.** Calculated reaction free energies (Δ$G$, eV) of OER elementary steps, OER overpotentials ($\eta_{OER}$, V), and potential-determining steps (PDSs) for $Co_2$-$C_{6-x}N_x$ DACs at 0 $V_{RHE}$ under constant-potential conditions.

| | $\Delta G_a$ | $\Delta G_b$ | $\Delta G_c$ | $\Delta G_d$ | $\eta_{OER}$ | PDS |
|---|---|---|---|---|---|---|
| $N_6$ | 1.47 | 1.80 | 1.06 | 0.59 | 0.57 | *OH → *O |
| $CN_5$-1 | 1.50 | 1.65 | 1.20 | 0.57 | 0.42 | *OH → *O |
| $C_2N_4$-12 | 1.28 | 1.81 | 1.11 | 0.72 | 0.58 | *OH → *O |
| $C_2N_4$-13 | 1.44 | 1.65 | 1.21 | 0.63 | 0.42 | *OH → *O |
| $C_2N_4$-14 | 1.32 | 1.61 | 1.31 | 0.68 | 0.38 | *OH → *O |
| $C_2N_4$-15 | 1.29 | 1.58 | 1.27 | 0.78 | 0.35 | *OH → *O |
| $C_2N_4$-16 | 1.44 | 1.65 | 1.21 | 0.62 | 0.42 | *OH → *O |
| $C_2N_4$-25 | 1.06 | 1.72 | 1.18 | 0.97 | 0.49 | *OH → *O |
| $C_3N_3$-123 | 1.04 | 1.66 | 1.27 | 0.94 | 0.43 | *OH → *O |
| $C_3N_3$-124 | 0.95 | 1.49 | 1.42 | 1.05 | 0.26 | *OH → *O |
| $C_3N_3$-125 | 0.82 | 1.38 | 1.50 | 1.22 | 0.27 | *O → *OOH |
| $C_3N_3$-126 | 1.04 | 1.65 | 1.29 | 0.94 | 0.42 | *OH → *O |
| $C_3N_3$-134 | 1.17 | 1.56 | 1.35 | 0.84 | 0.33 | *OH → *O |
| $C_3N_3$-135 | 1.07 | 1.52 | 1.33 | 0.99 | 0.29 | *OH → *O |
| $C_4N_2$-12 | 0.97 | 1.46 | 1.48 | 1.01 | 0.25 | *O → *OOH |
| $C_4N_2$-14 | 0.82 | 1.36 | 1.56 | 1.18 | 0.33 | *O → *OOH |
| $C_4N_2$-15 | 0.95 | 1.40 | 1.53 | 1.05 | 0.30 | *O → *OOH |
| $C_4N_2$-25 | 1.02 | 1.50 | 1.42 | 0.98 | 0.27 | *OH → *O |

**Table S24.** Hydrogen adsorption free energies($\Delta G_{*H}$) on $CoTM_2$-$N_6$ DACs ($TM_2$ = Ni, Cu and Co) at 0 $V_{RHE}$ under constant-potential conditions.

| Configurations | N1 | N2 | N3 | Co | $TM_2$ |
|---|---|---|---|---|---|
| CoNi-$N_6$ | 0.94 | 0.52 | 0.95 | 0.62 | 1.50 |
| CoCu-$N_6$ | 0.73 | 0.47 | 0.93 | 0.51 | 1.94 |
| $Co_2$-$N_6$ | 0.95 | 0.68 | -- | 0.75 | -- |

**Table S25.** Hydrogen adsorption free energies($\Delta G_{*H}$) on CoTM$_2$-C$_{6-x}$N$_x$ DACs (TM$_2$ = Ni, Cu and Co) at 0 V$_{RHE}$ under constant-potential conditions; $^a$ and $^b$ represent H adsorption at the X–TM bridge site and the top site (X = C), respectively.

| | | X1 | X2 | X3 | X4 | X5 | X6 | Co | TM$_2$ |
|---|---|---|---|---|---|---|---|---|---|
| CoNi | $C_3N_3$-124 | 0.08$^a$ | 0.26$^b$ | -- | 0.02$^a$ | -- | -- | -- | -- |
| | $C_3N_3$-134 | 0.12$^a$ | -- | 0.31$^b$ | 0.09$^a$ | -- | -- | -- | -- |
| | $C_3N_3$-156 | 0.14$^a$ | -- | -- | -- | 0.19$^b$ | 0.37$^b$ | -- | -- |
| | $C_4N_2$-16 | -- | 0.01$^b$ | 0.08$^a$ | -0.15$^a$ | 0.11$^b$ | -- | -- | -- |
| | $C_4N_2$-23 | -0.18$^a$ | -- | -- | -0.15$^a$ | 0.20$^b$ | 0.19$^a$ | -- | -- |
| | $C_4N_2$-25 | 0.03$^a$ | -- | 0.12$^b$ | -- | -- | -- | -- | -- |
| | $C_4N_2$-26 | -0.20$^a$ | -- | 0.13$^a$ | -0.12$^a$ | 0.25$^b$ | -- | -- | -- |
| | $C_4N_2$-36 | -0.20$^a$ | 0.13$^b$ | -- | -- | -- | -- | -- | -- |
| | $C_5N$-2 | -0.29$^a$ | -- | -0.11$^a$ | -0.25$^a$ | 0.14$^b$ | -0.06$^a$ | -- | -- |
| CoCu | $CN_5$-3 | -- | -- | 0.19$^b$ | -- | -- | -- | 0.53$^b$ | -- |
| | $C_2N_4$-15 | 0.10$^a$ | -- | -- | -- | -0.24$^b$ | -- | -- | -- |
| | $C_3N_3$-123 | 0.31$^a$ | 0.23$^b$ | 0.22$^b$ | -- | -- | -- | -- | -- |
| | $C_3N_3$-136 | 0.15$^a$ | -- | 0.17$^b$ | -- | -- | 0.13$^b$ | -- | -- |
| | $C_3N_3$-156 | 0.19$^a$ | -- | -- | -- | -0.08$^b$ | 0.24$^b$ | -- | -- |
| | $C_4N_2$-26 | 0.08$^a$ | -- | 0.24$^b$ | 0.19$^a$ | 0.19$^b$ | -- | -- | -- |
| | $C_4N_2$-36 | 0.02$^a$ | 0.16$^b$ | -- | -- | -- | -- | -- | -- |
| | $C_5N$-3 | -0.13$^a$ | 0.09$^b$ | -- | -0.19$^a$ | -0.19$^b$ | 0.22$^b$ | -- | -- |
| Co$_2$ | $C_3N_3$-124 | -0.20$^a$ | 0.33$^b$ | -- | -0.22$^a$ | -- | -- | -- | 0.31$^b$ |
| | $C_3N_3$-125 | 0.02$^a$ | 0.34$^b$ | -- | -- | 0.32$^b$ | -- | -- | 0.31$^b$ |
| | $C_3N_3$-135 | -0.12$^a$ | -- | -- | -- | 0.24$^b$ | -- | -- | -- |
| | $C_4N_2$-12 | -- | -- | -0.22$^a$ | -0.06$^a$ | 0.23$^b$ | -0.21$^a$ | -- | -- |
| | $C_4N_2$-15 | -- | 0.20$^b$ | -0.19$^a$ | -0.12$^a$ | -- | -0.25$^a$ | -- | -- |
| | $C_4N_2$-25 | -0.14$^a$ | -- | -- | -- | -- | -- | -- | -- |